\documentclass[reqno,twoside,11pt]{amsart}
\usepackage{amsmath}
\usepackage{amsfonts}
\usepackage{amssymb}
\usepackage{verbatim}

\IfFileExists{myowntimes.sty}{\usepackage{myowntimes}}
	{\usepackage{times}\usepackage{mathrsfs}}
\renewcommand{\slshape}{\itshape}
\renewcommand{\sffamily}{\rmfamily}

\def\version{September 8, 2000}
\newfont{\bfit}{cmbxti10 scaled 1200}
\newtheoremstyle{thm}{1.5ex}{1.5ex}{\itshape\rmfamily}{}
{\bfseries\rmfamily}{}{2ex}{}

\newtheoremstyle{def}{1.5ex}{1.5ex}{\rmfamily}{}
{\bfseries\rmfamily}{}{2ex}{}

\newtheoremstyle{rem}{1.3ex}{1.3ex}{\rmfamily}{}
{\itshape}
{} {1.5ex}{}

\newenvironment{proofsect}[1]
{\vskip0.1cm\noindent{\rmfamily\itshape#1.}}{\qed\vspace{0.15cm}}

\theoremstyle{thm}
\newtheorem{theorem}{Theorem}[section]
\newtheorem{lemma}[theorem]{Lemma}
\newtheorem{proposition}[theorem]{Proposition}

\newtheorem*{Main Theorem}{Main Theorem.}

\theoremstyle{rem}
\newtheorem{remark}{{\itshape Remark}}[]

\numberwithin{equation}{section}

\renewcommand{\section}{\secdef\sct\sect}
\newcommand{\sct}[2][default]{\refstepcounter{section}
\addcontentsline{toc}{section}
{{\tocsection {}{\thesection}{\!\!\!\!#1\dotfill}}{}}
\vspace{0.7cm}
\centerline{ 
\scshape\arabic{section}.\ #1} \nopagebreak \vspace{0.2cm}}
\newcommand{\sect}[1]{
\vspace{0.4cm} \centerline{\large\scshape\rmfamily #1}
\vspace{0.2cm}}

\renewcommand{\subsection}{\secdef\subsct\sbsect}
\newcommand{\subsct}[2][default]{\refstepcounter{subsection}
\addcontentsline{toc}{subsection}
{{\tocsection{\!\!}{\hspace{1.2em}\thesubsection}{\!\!\!\!#1\dotfill}}{}}
\nopagebreak\vspace{0.45\baselineskip} {\flushleft\bf
\arabic{section}.\arabic{subsection}~\bf #1.~}
\\*[3mm]\noindent
\nopagebreak}
\newcommand{\sbsect}[1]{\vspace{0.1cm}\noindent
\textbf{#1.~}\vspace{0.1cm}}

\renewcommand{\subsubsection}{%
\secdef \subsubsect\sbsbsect}
\newcommand{\subsubsect}[2][default]{%
\refstepcounter{subsubsection} 
\addcontentsline{toc}{subsubsection}{{\tocsection{\!\!}
{\hspace{3.05em}\thesubsubsection}{\!\!\!\!#1\dotfill}}{}}
\nopagebreak
\vspace{0.15\baselineskip} \nopagebreak {\flushleft\rmfamily
\itshape\arabic{section}.\arabic{subsection}.\arabic{subsubsection}
\ \rmfamily #1\/.}\ }
\newcommand{\sbsbsect}[1]{\vspace{0.1cm}\noindent
\rmfamily \itshape
\arabic{section}.\arabic{subsection}.\arabic{subsubsection} \
\sffamily #1\/.\ }

 \newcommand{\eps}{\varepsilon}
 \newcommand{\supp}{{\operatorname {supp}\,}}
 \newcommand{\prob}{{\operatorname {Prob}}}
 
 \newcommand{\dist}{{\operatorname {dist}}}

 \newcommand{\esssup}{{\operatorname {esssup}\,}}
 \newcommand{\const}{{\operatorname {const.}\,}}
 
 \renewcommand{\H}{\widetilde H}
 \newcommand{\HH}{{\mathbb H}}
 \newcommand{\R}{\mathbb{R}}
 \newcommand{\N}{\mathbb{N}}
 
 \renewcommand{\P}{\mathbb{P}}
 \newcommand{\Z}{\mathbb{Z}}
 \newcommand{\bdot}{\,\cdot\,}

 \newcommand{\E}{\mathbb{E}}
 \newcommand{\1}{{\sf 1}}

 \newcommand{\F}{{\mathcal F}}
 
 \newcommand{\skrih}{{\mathcal H}}
 \newcommand{\I}{{\mathcal I}}
 
 \renewcommand{\L}{{\mathcal L}}

\newcommand{\DeltaD}{\Delta^{\mkern-2.3mu\operatorname{d}}\mkern0.2mu}
\newcommand{\lambdaD}{\lambda^{\operatorname{d}}}
\newcommand{\nablaD}{\nabla}

\begin{document}


\title[Parabolic Anderson model]
{\Large Long-time tails in the parabolic Anderson
\\ model with bounded potential}

\author[Marek Biskup and Wolfgang K\"onig]{}
\maketitle

\thispagestyle{empty}
\vspace{0.2cm}
\centerline{\sc Marek Biskup$^1$\/
and Wolfgang K\"onig\small$^2$}
\vspace{0.8cm}
\centerline
{\em $^1$Microsoft Research, One Microsoft Way,
Redmond WA 98052, U.S.A.}
\centerline{\small and}
\centerline{\em $^2$Fachbereich Mathematik~MA7-5,
Technische Universit\"at Berlin,}
\centerline{\em
Stra\ss e des 17.~Juni~136, 10623 Berlin, Germany}
\vspace{0.5cm}
\centerline{\small(\version)}
\vspace{0.5cm}

\begin{quote}
{\footnotesize {\bf Abstract:} We consider the parabolic Anderson 
problem $\partial_t u=\kappa\Delta u+\xi u$ on $(0,\infty)\times 
\Z^d$ with random i.i.d.\ potential $\xi=(\xi(z))_{z\in\Z^d}$ and 
the initial condition $u(0,\cdot)\equiv1$. Our main assumption is 
that $\esssup\xi(0)=0$. Depending on the thickness of the 
distribution $\prob(\xi(0)\in\cdot)$ close to its essential 
supremum, we identify both the asymptotics of the moments of 
$u(t,0)$ and the almost-sure asymptotics of $u(t,0)$ as 
$t\to\infty$ in terms of variational problems. As a by-product, 
we establish Lifshitz tails for the random Schr\"odinger operator 
$-\kappa\Delta-\xi$ at the bottom of its spectrum. In our class 
of $\xi$~distributions, the Lifshitz exponent ranges from $d/2$ 
to $\infty$; the power law is typically accompanied by 
lower-order corrections.}
\end{quote}

\vfill
\begin{tabular}{lp{13cm}}
\multicolumn{2}{l}
\\ \footnotesize  almost-sure asymptotics, large deviations,
Dirichlet eigenvalues, percolation.\\
\end{tabular}

\eject

\section{Introduction and statement of results}

\subsection{Model and motivation}
In recent years, systems with {\it a priori\/} disorder have 
become one of the central objects of study in both probability 
theory and mathematical physics. Two of the pending open problems 
are the behavior of the simple random walk in random environment 
on the side of probability theory and understanding of the 
spectral properties of the so-called Anderson Hamiltonian on the 
side of (mathematical) solid state physics. The parabolic 
Anderson model studied in this paper encompasses various features 
of both aforementioned problems and thus provides a close link 
between the two seemingly rather remote areas. In particular, 
long-time tails in the parabolic model are intimately connected 
with the mass distribution of the spectral measure at the bottom 
of the spectrum for a class of Anderson Hamiltonians, and with 
the asymptotic scaling behavior of the random walk in random 
environment.

The parabolic Anderson model is the Euclidean-time (or diffusion)
version of the Schr\"odinger equation with a random potential.
More precisely, the name refers to the initial problem
\begin{equation}\label{Anderson}
\begin{array}{rcll}
\displaystyle
\partial_t \,u(t,z)\!\!\! &=&\!\!\!\kappa \DeltaD u(t,z)+\xi(z)
u(t,z),\qquad &(t,z)\in(0,
\infty)\times \Z^d,\\
u(0,z)\!\!\!&=&\!\!\!1,&z\in\Z^d,
\end{array}
\end{equation}
where $\partial_t$ is the time derivative,
$u\colon[0,\infty)\times\Z^d\to[0,\infty)$ is a function,
$\kappa>0$ is a diffusion constant, $\DeltaD$ is the discrete
Laplacian $[\DeltaD f](z)= \sum_{y\sim z}(f(y)-f(z))$ [here 
$y\sim z$ denotes that $y$ and $z$ are nearest neighbors], and 
$\xi=(\xi(z))_{z\in\Z^d}$ is a random i.i.d.\ potential. Let us 
use $\langle\bdot\rangle$ to denote the expectation with respect 
to  $\xi$ and let $\prob(\cdot)$ denote the corresponding 
probability measure. The main subject of our interest concerning 
\eqref{Anderson} is the large time behavior of the $p$-th moment 
$\langle u(t,0)^p\rangle$ for all $p>0$ and the almost-sure 
asymptotics of $u(t,0)$.

The quantity $u(t,z)$ can be interpreted as the expected total 
mass at time $t$ carried by a particle placed at time 0 at  site 
$z$ with a unit mass on it. The particle diffuses on $\Z^d$ like 
a simple random walk with generator $\kappa\DeltaD$; when present 
at site $x$, its mass is increased/decreased by an infinitesimal 
amount at rate $\pm\xi(x)\vee0$. Of particular interest is the 
phenomenon of {\it intermittency}: The total mass at time~$t$ 
comes mainly from passing through certain small $t$-dependent 
regions, the ``relevant islands,'' where the potential $\xi$ is 
large and of particular preferred shape. Intermittency is 
reflected (and sometimes defined) by a comparison of the 
asymptotics of $\langle u(t,0)^p\rangle^{1/p}$ for different $p$ 
and/or by a comparison of the growths of $\langle u(t,0)\rangle$ 
and $u(t,0)$, see also Remarks~\ref{intermitt1} and 
\ref{intermitt2} below. For general aspects of intermittency see 
G\"artner and Molchanov~\cite{GM90} and the monograph of Carmona 
and Molchanov~\cite{CM94}.

\subsection{Assumptions\label{assump}}
Since the time evolution in \eqref{Anderson} is driven by the
operator $\kappa\DeltaD+\xi$, it is clear that both large $t$
asymptotics of $u(t,0)$ are determined by the upper tails of the
random variable~$\xi(0)$. 
Our principal assumption  is that the
support of $\xi(0)$ is bounded from above. As
then follows by applying a criterion derived in G\"artner and
Molchanov \cite{GM90}, there is a unique non-negative solution to
\eqref{Anderson} for almost all $\xi$. Moreover, since
$\xi(\cdot)\to\xi(\cdot)+a$ is compensated by $u(t,\cdot)\to
e^{at}u(t,\cdot)$ in \eqref{Anderson}, we assume without loss of
generality that $\xi(0)$ is a non-degenerate random variable with
\begin{equation}
\label{mainass}
\esssup\xi(0)=0.
\end{equation}
Hence, our potential $\xi$ is non-positive throughout $\Z^d$, 
i.e., every lattice site $x$ is either neutral ($\xi(x)=0$) or a 
``soft trap'' ($-\infty<\xi(x)<0$) or a ``hard trap'' 
($\xi(x)=-\infty$). Furthermore, $\xi(x)$ exceeds any negative 
value with positive probability. Note that {\it a priori\/} we do 
not exclude hard traps, but some restrictions to the size of 
$\prob(\xi(0)=-\infty)$ have to be imposed in order to have an 
interesting almost-sure asymptotics (see Theorem~\ref{asasy}). 
The important special case of ``Bernoulli traps,'' where the 
potential attains only the values 0 and $-\infty$, has already 
extensively been studied by, e.g., Donsker and Varadhan 
\cite{DV79}, Antal \cite{A95}, and in a continuous analogue by 
Sznitman~\cite{S98}.

As we have indicated above, our results will prominently depend 
on the asymptotics of $\prob(\xi(0)>-x)$ as $x\downarrow 0$. 
Actually, they turn out to depend on two parameters $A\in 
(0,\infty)$ and $\gamma\in[0,1)$ only, which appear as follows:
\begin{equation}\label{mainex}
\prob\bigl(\xi(0)>-x\bigr)=\exp\left\{-A\,
x^{-\frac{\gamma}{1-\gamma}+o(1)}\right\}, \qquad x\downarrow 0.
\end{equation}
The reader should keep \eqref{mainex} in mind as the main 
representative of the distributions we are considering. The case 
$\gamma=0$ contains the above mentioned special case of 
``Bernoulli traps.''

However, our precise assumption on the thickness of 
$\prob(\xi(0)\in\cdot)$ at zero will be more technical. As turns 
out to be more convenient for our proofs, we describe the upper 
tail of $\prob(\xi(0)\in\cdot)$ in terms of scaling properties of 
the cumulant generating function
\begin{equation}\label{Hdef}
H(\ell)=\log\langle e^{\ell\xi(0)}\rangle,\qquad \ell\ge0.
\end{equation}
The reason is that $H$ naturally appears once expectation with
respect to $\xi$ is taken on the Feynman-Kac representation of
$u(t,0)$, see e.g.\ formula \eqref{lowbound}. Note that $H$ is
convex and, by \eqref{mainass}, decreasing and strictly negative
on $(0,\infty)$.

\vspace{0.2cm}
\noindent{\bf Scaling Assumption.}
{\it We assume that there is a non-decreasing function
$t\mapsto\alpha_t\in(0,\infty)$ and a function $\widetilde
H\colon[0,\infty)\to(-\infty,0]$, $\widetilde H\not\equiv 0$,
such that}
\begin{equation}
\label{Hscaling}
\lim_{t\to\infty}\frac{\alpha_t^{d+2}}t
H\left(\frac{t}{\alpha_t^d}\, y\right) = \widetilde H(y),\qquad
y\ge0,
\end{equation}
{\it uniformly on compact sets in $(0,\infty)$.} \vspace{1.5mm}
 
Informally and intuitively, the scale function $\alpha_t$ admits 
the interpretation as the asymptotic diameter of the ``relevant 
islands'' from which the main contribution to the expected total 
mass $\langle u(t,0)\rangle$ comes; see also 
Subsection~\ref{Heuristics}. The choice of the scaling ratios 
$\alpha_t^{d+2}/t$ and $t/\alpha_t^d$ in \eqref{mainass} is 
dictated by matching two large-deviation scales: one (roughly) 
for the range of the simple random walk, the other for the size 
of the field $\xi$, see Subsection~\ref{Heuristics}. 

\begin{remark}
The finiteness and non-triviality of $\widetilde H$
necessitate that $t/\alpha_t^d\to\infty$ and
$\alpha_t=O(t^{1/(d+2)}$). In the asymptotic sense,
\eqref{Hscaling} and non-triviality of $\widetilde H$ determine
the pair $(\alpha_t,\widetilde H)$ uniquely up to a constant
multiple resp.\ scaling. Indeed, if
$(\widehat\alpha_t,\widehat{H})$ is another pair satisfying the
Scaling Assumption then, necessarily,
$\widehat\alpha_t/\alpha_t\to c\not=0,\infty$ and
$\widehat{H}(\cdot)=c^{d+2}\widetilde H(\cdot/c^d)$. Moreover, if
$t\mapsto\widehat\alpha_t$ is a positive function with
$\widehat\alpha_t/\alpha_t\to0$, then the limit in
\eqref{Hscaling} gives $\widehat{H}\equiv0$. Similarly, if
$\widehat\alpha_t/\alpha_t\to\infty$, then $\widehat{H}
\equiv-\infty$. These assertions follow directly from convexity of
$H$ (see also Subsection~\ref{sub-powlaw}).
\end{remark}

Our Scaling Assumption should be viewed as a more general form of 
\eqref{mainex} that is better adapted to our proofs.  
Remarkably, it actually  constrains
the form of possible $\widetilde H$ to a two-parameter family and
forces the scale function $\alpha_t$ to be regularly varying.
The following claim is proved in Subsection~\ref{sub-powlaw}.

\vbox{
\begin{proposition}
\label{power-law}
Suppose that \eqref{mainass} and the Scaling
Assumption hold. Then
\begin{equation}
\label{perfscalforH}
\widetilde H(y)=\widetilde H(1) y^\gamma,\qquad y>0,
\end{equation}
for some $\gamma\in[0,1]$.
Moreover,
\begin{equation}
\label{scalingexp1}
\lim_{t\to\infty}\frac{\alpha_{pt}}{\alpha_{t}}
=p^{\nu}\quad\mbox{ for all }\, p>0\/,\mbox{ and }\quad
\lim_{t\to\infty}\frac{\log \alpha_t}{\log t}=\nu,
\end{equation}
where
\begin{equation}\label{nudef}
\nu=\frac{1-\gamma}{d+2-d\gamma}\in \bigl[0,\textstyle\frac
1{d+2}\bigr].
\end{equation}
\end{proposition}
}

\begin{remark}
As is seen from \eqref{mainex}, each value $\gamma\in[0,1)$ can be
attained. Note that, despite the simplicity of possible
$\widetilde H$, the richness of the class of all
$\xi$~distributions persists in the scaling behavior of
$\alpha_t=t^{\nu+o(1)}$. For instance, the case $\gamma=0$
includes both distributions with an atom at $0$ and those with no
atom but with a density $\rho$ (w.r.t.\ the Lebesgue measure)
having the asymptotic behavior $\rho(x)\sim (-x)^\sigma$
$(x\uparrow 0)$ for a $\sigma>-1$. It is easy to find that
$\alpha_t=t^{1/(d+2)}$ [and $\widetilde H(1)=\log\prob(\xi(0)=0)$]
in the first case while $\alpha_t=(t/\log t)^{1/(d+2)}$ in the
second one. Yet thinner a tail has $\rho(x)\sim
\exp(-\log^\tau|x|^{-1})$ with $\tau>1$, for which we find
$\alpha_t=(t/\log^{\tau} t)^{1/(d+2)}$. Similar examples exist for
any $\gamma\in[0,1)$.
\end{remark}

Proposition~\ref{power-law} leads us to the following useful
concept:

\vspace{0.2cm} \noindent {\bf Definition.\ } {\it  Given a
$\gamma\in[0,1]$, we say that $H$ is\/ {\rm in the
$\gamma$-class\/}, if \eqref{mainass} holds and there is a
function $t\mapsto\alpha_t$ such that $(H,\alpha_t)$ satisfies the
Scaling Assumption and the limiting  $\widetilde H$ is homogeneous
with exponent $\gamma$, as~in~\eqref{perfscalforH}.}
\vspace{0.2cm}

Throughout the remainder of this paper, we restrict ourselves to
the case $\gamma<1$. The case $\gamma=1$ is qualitatively
different from that of $\gamma<1$; for more explanation see
Subsections~\ref{gamma=1} and~\ref{onourmethod}.

The rest of this paper is organized as follows. In the remainder
of this section we state our results (Theorems~\ref{momasy}
and~\ref{asasy}) on the moment and almost-sure asymptotics of
$u(t,0)$ and on Lifshitz tails of the Schr\"odinger operator
$-\kappa\DeltaD-\xi$ (Theorem~\ref{Lifshitz}). The next section
contains heuristic explanation of the proofs, discussion of the
case $\gamma=1$ in \eqref{mainex}, some literature remarks, and a
list of open problems. Section~\ref{prepare} contains necessary
definitions and proofs of some technical claims (in particular,
Proposition~\ref{power-law}). The
proofs of our main results (Theorems~\ref{momasy}
and~\ref{asasy}) come in Sections~\ref{proofmom}
and~\ref{asproof}.

\subsection{Main results}
\vspace{-7mm}
\subsubsection{Fundamental objects}\label{objects}
First we introduce some objects needed for the definition of the 
quantity $\chi$ which is basic for all our results. An 
uninterested reader may consider skipping these definitions and 
passing directly to Subsection~\ref{mom}.

\smallskip
$\bullet$ {\sl Function spaces}: Define
\begin{equation}
\label{functionclasses}
{\mathcal F}=\left\{f\in C_{\rm c}(\R^d,[0,\infty))\colon\Vert
f\Vert_1=1\right\},
\end{equation}
and for $R>0$, let ${\mathcal F}_R$ be set of $f\in{\mathcal F}$
with support in $[-R,R]^d$. By $C^+(R)$ (resp.\ $C^-(R)$) we
denote the set of continuous functions $[-R,R]^d\to [0,\infty)$
(resp.\ $[-R,R]^d\to (-\infty,0]$). Note that functions
in ${\mathcal F}_R$ vanish at the boundary of $[-R,R]^d$, while
those in $C^\pm(R)$ may not.

\smallskip
$\bullet$
{\sl Functionals}: Let ${\mathcal I}\colon{\mathcal F}
\to[0,\infty]$ be the Donsker-Varadhan rate functional
\begin{equation}
\label{Idef}
{\mathcal I}(f)=\begin{cases} \kappa\bigl\Vert(-\Delta)^{\frac
12}\sqrt{f} \bigr\Vert_2^2 &\mbox{if } \sqrt{f}\in{\mathcal
D}\bigl((-\Delta)^{\frac 12}\bigr),\\ \infty&\mbox{otherwise,}
\end{cases}
\end{equation}
where $\Delta$ is the Laplace operator on $L^2(\R^d)$ (defined as
a self-adjoint extension of $\sum_i(\partial^2/\partial x_i^2)$
from, e.g., the Schwarz class on $\R^d$) and ${\mathcal
D}((-\Delta)^{1/2})$ denotes the domain of its square root. Note
that ${\mathcal I}(f)$ is nothing but the Dirichlet form of the
Laplacian evaluated at $f^{1/2}$.

For $R>0$ we define the functional ${\mathcal H}_R
\colon C^+(R)\to (-\infty,0]$ by putting
\begin{equation}
\label{calHdef}
{\mathcal H}_R(f)=\int_{[-R,R]^d}\widetilde H\bigl(f(x)\bigr)\,dx.
\end{equation}
Note that for $H$ in the $\gamma$-class, ${\mathcal
H}_R(f)\!=\!\widetilde H(1)\int f(x)^\gamma dx$, with the
interpretation ${\mathcal H}_R(f)\!=\!\widetilde H(1)|\supp f|$
when $\gamma=0$. Here $|\cdot|$ denotes the Lebesgue measure.

\smallskip
$\bullet$
{\sl Legendre transforms}:
Let ${\mathcal L}_R\colon
C^-(R)\to[0,\infty]$ be the Legendre transform of ${\mathcal
H}_R$,
\begin{equation}\label{calLdef}
{\mathcal L}_R(\psi)=\sup\bigl\{(f,\psi)-{\mathcal
H}_R(f)\colon f\in C^+(R),\,\supp f\subset\supp\psi\bigr\},
\end{equation}
where we used the shorthand notation
$(f,\psi)=\int f(x)\psi(x)\,dx$. If $H$ is in the
$\gamma$-class, we get ${\mathcal L}_R(\psi)=\const \int
|\psi(x)|^{-\frac\gamma{1-\gamma}}\,dx$ for $\gamma\in(0,1)$ and
${\mathcal L}_R(\psi)=-\widetilde H(1)\,|\supp \psi|$ for $\gamma=0$.

For any potential $\psi\in C^-(R)$, we also need the principal
(i.e., the largest) eigenvalue of the operator $\kappa\Delta +
\psi$ on $L^2([-R,R]^d)$ with Dirichlet boundary conditions,
expressed either as the Legendre transform of $\I$ or in terms of
the Rayleigh-Ritz principle:
\begin{equation}
\begin{aligned}
\label{lambdadef}
\lambda_R(\psi)&=\sup\bigl\{(f,\psi)-{\mathcal I}(f)\colon f\in
{\mathcal F}_R,\,\supp f\subset \supp\psi\bigr\}\\
&=\sup\bigl\{(\psi, g^2)-\kappa\|\nabla g\|_2^2\colon g\in C_{\rm
c}^\infty(\supp\psi,\R),\|g\|_2 =1\bigr\},
\end{aligned}
\end{equation}
with the interpretation $\lambda_R(0)=-\infty$.

\smallskip
$\bullet$ {\sl Variational principles}: Here is the main quantity
of this subsection:
\begin{align}
\label{chidef}
\chi&=\inf_{R>0}\inf
\bigl\{{\mathcal I}(f)-{\mathcal H}_R(f)\colon f
\in{\mathcal F}_R\bigr\}\\
\label{chiident}
&=\inf_{R>0}
\inf\bigl\{{\mathcal L}_R
(\psi)-\lambda_R(\psi)\colon \psi\in C^-(R)\bigr\}.
\end{align}
where \eqref{chiident} is obtained from \eqref{chidef} by
inserting \eqref{calLdef} and the second line in
\eqref{lambdadef}. Note that $\chi$ depends on $\gamma$ and the
constant $\widetilde H(1)$.
\vspace{2mm}

\subsubsection{Moment asymptotics}
\label{mom}
We proceed by describing the logarithmic asymptotics of the
$p$-th moment of $u(t,0)$; for
the proof see Section~\ref{proofmom}.

\begin{theorem}\label{momasy}
Suppose that \eqref{mainass} and the Scaling Assumption hold. Let
$H$ be in the $\gamma$-class for some $\gamma\in[0,1)$. Then
$\chi\in(0,\infty)$ and
\begin{equation}\label{chi}
\lim_{t\to\infty}\frac{\alpha_{pt}^2} {pt}\log
\bigl\langle u(t,0)^p\bigr\rangle=-\chi,
\end{equation}
for every $p\in(0,\infty)$.
\end{theorem}

\begin{remark}
Both formulas \eqref{chidef} and \eqref{chiident} arise in
well-known large-deviation statements: the former for an
exponential functional of Brownian occupation times, the latter
for the principal eigenvalue for a scaled version of the field
$\xi$. Our  proof pursues the route leading to \eqref{chidef}; an
approach based on the second formula is heuristically explained in
Subsection~\ref{MA}.
\end{remark}

\begin{remark}\label{intermitt1}
Formula \eqref{chi}, together with the results of
Proposition~\ref{power-law}, imply that
\begin{equation}
\label{intermittency}
\lim_{t\to\infty}\frac{\alpha_{t}^2} {t}\log\frac{
\langle u(t,0)^p\rangle^{1/p}}{
\langle u(t,0)^q\rangle^{1/q}}=\chi \bigl(q^{-2\nu}-p^{-2\nu}\bigr),
\qquad p,q\in(0,\infty),
\end{equation}
whenever $H$ is in the $\gamma$-class, where $\nu>0$ is as in
\eqref{nudef}. In particular, $\langle u(t,0)^p\rangle$ for $p>1$
decays much slower than $\langle u(t,0)\rangle^p$. This is one
widely used manifestation of intermittency.
\end{remark}

\subsubsection{Lifshitz tails}
Based on Theorem~\ref{momasy}, we can compute the asymptotics of
the so-called\/ {\it integrated density of states\/} (IDS) of the
operator $-\kappa\DeltaD-\xi$ on the right-hand side of
\eqref{Anderson}, at the bottom of its spectrum. Below we define
the IDS and list some of its basic properties. For a
comprehensive treatment and proofs we refer to the book by
Carmona and Lacroix~\cite{CL90}.

The IDS is defined as follows: Let $R>0$ and let us consider the 
operator ${\mathfrak H}_R=-\kappa\DeltaD-\xi$ in 
$[-R,R]^d\cap\{x\in{\mathbb Z}^d\colon\xi(x)>-\infty\}$ with 
Dirichlet boundary conditions. Clearly, ${\mathfrak H}_R$ has a 
finite number of eigenvalues that we denote $E_k$, so it is 
meaningful to consider the quantity
\begin{equation}
\label{numberofstates}
N_R(E)=\#\{ k\colon E_k\le E\},\qquad E\in\R.
\end{equation}
The integrated density of states is then the limit
\begin{equation}
\label{IDSlimit}
n(E)=\lim_{R\to\infty}\frac{N_R(E)}{(2R)^d},
\end{equation}
giving $n(E)$ the interpretation as the number of energy levels
below $E$ per unit volume. The limit exists and is almost surely
constant, as can be proved using e.g. subadditivity.

It is clear that $E\mapsto n(E)$ is monotone and that $n(E)=0$
for all $E<0$, provided \eqref{mainass} is assumed. In the
1960's, based on heuristic arguments, Lifshitz postulated that
$n(E)$ behaves like $\exp(-\const E^{-\delta})$ as $E\downarrow
0$. This asymptotic form has been established rigorously in the
so called ``obstacle cases'' (see Subsection~\ref{relwork})
treated by Donsker and Varadhan \cite{DV79} and Sznitman
\cite{S98}, with $\delta=d/2$. Here we generalize this result to
our class of distributions with $\gamma<1$; however, in our cases
the power-law is typically supplemented with a lower-order
correction. The result can concisely be formulated in terms of
the inverse function of $t\mapsto\alpha_t$:

\begin{theorem}
\label{Lifshitz} Suppose that \eqref{mainass} and the Scaling
Assumption hold. Let $H$ be in the $\gamma$-class for some
$\gamma\in[0,1)$ and let $\alpha^{-1}$ be the inverse to the
scaling function $t\mapsto\alpha_t$. Then
\begin{equation}
\lim_{E\downarrow0}\,
\frac{\log n(E)}{E\alpha^{-1}(E^{-\frac12})}
=-\frac{2\nu}{1-2\nu}\bigl[(1-2\nu)\chi\bigr]^{-\frac1{2\nu}}
\end{equation}
where $\chi$ is as in \eqref{chidef} and $\nu$ is defined in
\eqref{nudef}.
\end{theorem}

Invoking \eqref{scalingexp1},
$E\alpha^{-1}(E^{-1/2})=E^{-1/\beta+o(1)}$ as $E\downarrow0$,
where
\begin{equation}
\label{beta}
\beta=\frac2{d+2\frac\gamma{1-\gamma}}=\frac{2\nu}{1-2\nu}\in
\left(0,\textstyle\frac 2d\right].
\end{equation}
In particular, $1/\beta$ is the Lifshitz exponent.
Theorem~\ref{Lifshitz} is proved in Subsection~\ref{pfofLifshitz}.

\subsubsection{Almost-sure asymptotics}
The almost-sure behavior of $u(t,0)$ depends strongly on whether
the origin belongs to a finite or infinite component of the set
${\mathcal C}=\{z\in\Z^d\colon\xi(z)>-\infty\}$. Indeed, if $0$ is
in a finite component of ${\mathcal C}$, then $u(t,0)$ decays
exponentially with $t$. Thus, in order to get a non-trivial
almost-sure behavior of $u(t,0)$ as $t\to\infty$, we need that
${\mathcal C}$ contains an infinite component ${\mathcal
C}_\infty$ and that $0\in{\mathcal C}_\infty$ occurs with a
non-zero probability. In $d\geq2$, this is guaranteed by requiring
that $\prob(\xi(0)>-\infty)$ exceed the percolation threshold
$p_{\rm c}(d)$ for site percolation on $\Z^d$. In $d=1$,
${\mathcal C}$ is percolating if and only if
$\prob(\xi(0)>-\infty)=1$; sufficient ``connectivity'' can be
ensured only under an extra condition on the {\it lower\/} tail of
$\xi(0)$.

Suppose, without loss of generality, that $t\mapsto t/\alpha_t^2$
is strictly increasing (recall that $\alpha_t=t^{\nu+o(1)}$ with
$\nu\le 1/3$). Then we can define another scale function $t\mapsto
b_t\in(0,\infty)$ by setting
\begin{equation}\label{btdef}
\frac{b_t}{\alpha_{b_t}^2} = \log t,\qquad t>0.
\end{equation}
(In other words, $b_t$ is the inverse function of $t\mapsto
t/\alpha_t^2$ evaluated at $\log t$.) Let
\begin{equation}
\label{chitildedef}
\widetilde\chi=-\sup_{R>0}\,\sup\left\{\lambda_R(\psi)\colon\psi\in
C^-(R),\, {\mathcal L}_R(\psi)\leq d\right\}.
\end{equation}
In our description of the almost sure asymptotics, the pair
$(\alpha_{b_t},\widetilde \chi)$ will play a role analogous to the
pair $(\alpha_t,\chi)$ in Theorem~\ref{momasy} [in particular,
$\alpha_{b_t}$ is the diameter of the ``islands'' in the
``$\xi$~landscape'' dominating the a.s.\  asymptotics of
$u(t,0)$]. It is clear from Proposition~\ref{power-law} that
\begin{equation}
\label{growthofb_t}
b_t=(\log t)^{\frac1{1-2\nu}+o(1)}\quad\text{ and }\quad
\alpha_{b_t}^2=\bigl(\log t\bigr)^{\beta+o(1)},\qquad t\to\infty,
\end{equation}
where $\beta$ is as in \eqref{beta}. It turns out that
$\widetilde\chi$ can be computed from $\chi$:

\begin{proposition}
\label{tildechiident}
Suppose that \eqref{mainass} and the Scaling Assumption hold. Let
$H$ be in the $\gamma$-class for some $\gamma\in[0,1)$. Let $\nu$
and $\beta$ be as in \eqref{scalingexp1} and \eqref{beta}. Then
$\widetilde\chi\in(0,\infty)$ and
\begin{equation}
\label{chichirel}
\widetilde\chi=\chi^{\frac 1{1-2\nu}} (1-2\nu)\left(\frac
{2\nu}d\right)^\beta,
\end{equation}
where $\chi$ and $\widetilde\chi$ are as in \eqref{chidef}
and \eqref{chitildedef}.
\end{proposition}

The proof of Proposition~\ref{tildechiident} is given in
Subsection~\ref{tildechiproof}. In the special case $\gamma=0$,
the relation \eqref{chichirel} can independently be verified by
inserting the explicit expressions for $\chi$ and $\widetilde\chi$
derived e.g.\ in Sznitman~\cite{S98}.

Our main result on the almost sure asymptotics reads as follows:

\begin{theorem}\label{asasy}
Suppose that \eqref{mainass} and the Scaling Assumption hold. Let
$H$ be in the $\gamma$-class for some $\gamma\in[0,1)$. In
$d\ge2$, let $\prob(\xi(0)>-\infty)>p_{\rm c}(d)$; in $d=1$, let
$\langle\log(-\xi(0)\vee 1)\rangle<\infty$. Then
\begin{equation}\label{asasyformula}
\lim_{t\to\infty}\frac {\alpha_{b_t}^2}t\log
u(t,0)=-\widetilde\chi \quad\qquad \prob(\,\cdot\,|0\in {\mathcal
C}_\infty)\text{\rm{}-almost surely}.
\end{equation}
\end{theorem}
\vspace{0.1cm}

Theorem~\ref{asasy} is proved in Section~\ref{asproof}; for a
heuristic derivation see Subsection~\ref{ASA}.

\begin{remark}\label{intermitt2}
From a comparison of the asymptotics in \eqref{chi} and in 
\eqref{asasyformula}, we obtain another manifestation of 
intermittency: The moments of $u(t,0)$ decay much slower than 
the $u(t,0)$ itself.
\end{remark}

Assuming that there is no critical site percolation in dimensions
$d\ge 2$, Theorem~\ref{asasy} and the arguments at the beginning
of this subsection give a complete description of possible
leading-order almost-sure asymptotics of $u(t,0)$. 

\begin{remark}
In $d=1$, there {\it is\/} site percolation at $p_c(1)=1$ which 
is the reason why an extra condition on the lower tail of 
$\prob(\xi(0)\in\cdot)$ needs to be assumed. If the lower tail 
is too heavy, i.e., if $\log(-\xi(0)\vee 1)$ is not integrable, 
then a {\it screening effect\/} occurs: The mass flow over large 
distances is hampered by regions of large negative field, which 
cannot be circumvented due to one-dimensional topology. As has 
recently been shown in Biskup and K\"onig \cite{BK00}, $u(t,0)$ 
decays faster than in the cases described in Theorem~\ref{asasy}. 
\end{remark} 

\section{Heuristics, literature remarks, and open problems}
\label{blahblah}
\subsection{Heuristic derivation\label{Heuristics}}
In our heuristics we use the interpretation of \eqref{Anderson} 
in terms of a particle {\it system\/} that randomly evolves in a 
random potential of traps: A particle at $z$ either jumps to its 
nearest neighbor at rate $\kappa$ or is killed at rate $-\xi(z)$. 
Then $u(t,0)$ is the total expected number of particles located 
at the origin at time~$t$, provided the initial configuration had 
exactly one particle at each lattice site.

It is clear from \eqref{mainass} that, by time $t$, the origin is
not likely to be reached by any particle from regions having
distance more than $t$ from the origin. If $u_t(t,0)$ is the
expected number of particles at the origin at time $t$ under the
constraint that none of the particles has ever been outside of
the box $Q_t=[-t,t]^d\cap\Z^d$, then this
should imply that
\begin{equation}\label{heu1}
u(t,0)\approx u_t(t,0).
\end{equation}
The particle system in the box $Q_t$ is driven by the operator
$\kappa\DeltaD+\xi$ on the right-hand side of \eqref{Anderson}
with zero boundary conditions on $\partial Q_t$ and the
leading-order behavior of $u_t$ should be governed by its
principal (i.e., the largest) eigenvalue $\lambda^{\rm d}_t(\xi)$
in the sense that
\begin{equation}\label{heu2}
u_t(t,0)\approx e^{t\lambda^{\rm d}_t(\xi)}.
\end{equation}
Based on \eqref{heu2}, we can give a plausible explanation
of our Theorems~\ref{momasy} and~\ref{asasy}.

\subsubsection{Moment asymptotics}
\label{MA}
Under the expectation with respect to $\xi$, there is a
possibility that $\langle u(t,0)\rangle$ will be dominated by a
set of $\xi$'s with exponentially small probability. But then the
decisive contribution to the average particle-number at zero may
come from much smaller a box than $Q_t$. Let $R\alpha_t$ denote
the diameter of the purported box. Then we should have
\begin{equation}
\label{heu3}
\bigl\langle u_t(t,0)\bigr\rangle\approx \bigl\langle e^{t\lambda^{\rm
d}_{R\alpha_t}}\bigr\rangle.
\end{equation}
The proper choice of the scale function $\alpha_t$ is determined
by balancing the gain in $\lambda^{\rm d}_{R\alpha_t}(\xi)$ and
the loss due to taking $\xi$'s with exponentially small
probability. Introducing the scaled field
\begin{equation}\label{scaledfield}
\bar\xi_t(x)=\alpha_t^2\xi\bigl(\lfloor x\alpha_t\rfloor\bigr),
\end{equation}
the condition that these scales match for $\bar\xi_t\approx\psi\in
C^-(R)$ reads
\begin{equation}
\label{asymp} \log\prob(\bar\xi_t\approx \psi) \asymp
t\lambda^{\rm
d}_{R\alpha_t}\bigl(\alpha_t^{-2}\psi(\cdot\,\alpha_t^{-1})\bigr).
\end{equation}
By scaling properties of the continuous Laplace operator, the
right-hand side is approximately equal to $(t/\alpha_t^2)\lambda_R(\psi)$,
where $\lambda_R(\psi)$ is defined in \eqref{lambdadef}. On the
other hand, by our Scaling Assumption,
\begin{equation}\label{heuldp}
\log\prob(\bar\xi_t\approx \psi) \approx-\frac t{\alpha_t^2}{\mathcal
L}_R(\psi),
\end{equation}
i.e., we expect $\bar\xi_t$ to satisfy a large-deviation principle
with rate $t/\alpha_t^2$ and rate function ${\mathcal L}_R$. Then
the rates on both sides of \eqref{asymp} are identical and,
comparing also the prefactors, we have
\begin{equation}\label{heu4}
\bigl\langle e^{t\lambda_{R\alpha_t}^{\rm
d}}\1\{\bar\xi_t\approx\psi\}\bigr\rangle \approx
\exp\bigl\{\textstyle{ \frac t{\alpha_t^2}
[\lambda_R(\psi)-{\mathcal L}_R(\psi)]}\bigr\}.
\end{equation}
Now collect \eqref{heu1}, \eqref{heu3} and \eqref{heu4} and
maximize over $\psi\in C^-(R)$ and over $R>0$ to obtain formally
the statement on the moment asymptotics in Theorem~\ref{momasy}
for $p=1$. Note that, by the above heuristic argument, $\alpha_t$
is the spatial scale of the ``islands'' in the potential landscape
that are only relevant for the moments of $u(t,0)$.

\subsubsection{Almost-sure asymptotics}
\label{ASA} Based on the intuition developed for the moment
asymptotics, the decisive contribution to \eqref{heu2} should
come from some quite localized region in $Q_t$. Suppose this
region has size $\alpha_{b_t}$, where $b_t$ is some new running
time scale, and divide $Q_t$ regularly into boxes of diameter
$R\alpha_{b_t}$ (``microboxes'') with some  $R>0$. According to
\eqref{heuldp} with $t$ replaced by $b_t$, we have for any
$\psi\in C^-(R)$ with ${\mathcal L}_R(\psi)\le d$ that
\begin{equation}
\label{t^d}
\prob(\bar\xi_{b_t}\approx\psi)\approx
\exp\bigl\{\textstyle{-\frac{b_t}{\alpha_{b_t}^2}
{\mathcal L}_R(\psi)}\bigr\}
\geq e^{-d b_t/\alpha_{b_t}^2},
\end{equation}

Suppose that $b_t$ obeys \eqref{btdef}. Then the right-hand side of
\eqref{t^d} decays as fast as $t^{-d}$. Since there
are of order $t^d$ microboxes in $Q_t$, a Borel-Cantelli argument
implies that for any $\psi$ with ${\mathcal L}_R(\psi)< d$, there
will be a microbox in $Q_t$ where $\bar\xi_{b_t}\approx\psi$. As
before, $t\lambda_{R\alpha_{b_t}}^{\rm d}
(\psi(\cdot/\alpha_{b_t})/\alpha_{b_t}^{2})\approx
(t/\alpha_{b_t}^2)\lambda_R(\psi)$, and by optimizing over $\psi$,
any value smaller than $\widetilde\chi$ can be attained by
$\lambda_R(\psi)$ in some microbox in~$Q_t$.

This suggests that $u(t,\cdot)$ in the favorable microbox decays
as described by \eqref{asasyformula}. It remains to ensure, and
this is a non-trivial part of the argument, that the particles
that have survived in this microbox by time $t$ can always reach the
origin within a negligible portion of time $t$. This requires, in
particular, that sites $x$ with $\xi(x)>-\infty$ form an infinite cluster
containing the origin. If the connection between $0$ and the
microbox can be guaranteed, $u(t,0)$ should exhibit the same
leading-order decay, which is the essence of the claim in
Theorem~\ref{asasy}. Note that, as before, $\alpha_{b_t}$ is the
spatial scale of the islands relevant for the random variable
$u(t,0)$.

\subsection{The case $\boldsymbol{\gamma=1}$\label{gamma=1}}
In the boundary case $\gamma=1$ the relevant islands grow
(presumably) slower than any polynomial as $t\to\infty$ (i.e.,
$\alpha_t=t^{o(1)}$), and  $\widetilde H$ is linear. As a
consequence, the asymptotic expansions of  $\langle
u(t,0)^p\rangle$ and $u(t,0)$ itself start with a {\it
field-driven\/} term (i.e, a term independent of $\kappa$). In
particular, no variational problem is involved at the leading
order and no information about the ``typical'' configuration of
the fields is gained.

To understand which $\xi$ dominate the moments of $u(t,0)$ we have
to analyze the next-order term. This requires imposing an
additional assumption: We suppose the existence of a new scale
function $t\mapsto\vartheta_t$, with $\alpha_t=o(\vartheta_t)$,
such that
\begin{equation}
\label{Heta} \lim_{t\to\infty}\textstyle\frac
{\vartheta_t^{d+2}}t\Bigl[H\bigl(\textstyle\frac t{\vartheta_t^d}
y\bigr)- H\bigl(\textstyle\frac
t{\vartheta_t^d}\bigr)y\Bigr]=\widehat H(y)
\end{equation}
exists (and is not identically zero) locally uniformly in
$y\in(0,\infty)$. Analogous heuristic to that we used to explain
the main idea of Theorem~\ref{momasy} outputs the asymptotic
expansion of the first moment
\begin{equation}
\bigl\langle u(t,0)\bigr\rangle=\exp\Bigl[\vartheta_t^d
H\bigl(\textstyle\frac{t}{\vartheta_{t}^d}\bigr)-
(t/\vartheta_t^2) \bigl(\widehat\chi+o(1)\bigr)\Bigr],
\end{equation}
where $\widehat\chi$ is defined as in Subsection~\ref{objects} with
$\widetilde H$ replaced by $\widehat H$.

Similar scenario should occur for the almost-sure asymptotics.
Indeed, setting
\begin{equation}
\psi(x)=(\vartheta_t^{d}/t)H(t/\vartheta_t^d)+
\vartheta_t^{-2}\psi_\star(x/\vartheta_t)
\end{equation}
with some $\psi_\star\in C^-(R)$, formula \eqref{t^d} should be 
rewritten as $\prob(\xi\approx\psi)\approx 
\exp\{-(t/\vartheta_t^2) {\mathcal L}_R^\star(\psi_\star)\}$, 
where ${\mathcal L}_R^\star$ is defined by \eqref{calLdef} with 
$\widetilde H$ replaced by $\widehat H$. Let $b_t^\star$ solve 
for $s$ in $s/\vartheta_s^2=\log t$. By following the heuristic 
derivation of Theorem~\ref{asasy} (and, in particular, invoking 
the scaling and additivity of the continuum eigenvalue 
$\lambda_R(\psi)$, see Subsection~\ref{ASA}) we find that
\begin{equation}
u(t,0)=\exp\Bigl[(t\vartheta_{b_t^\star}^d/b_t^\star)
H(b_t^\star/\vartheta_{b_t^\star}^d) -(t/\vartheta_{b_t^\star}^2)
\bigl(\widehat\chi_\star+o(1)\bigr)\Bigr]
\end{equation}
should hold $\prob(\cdot|0\in{\mathcal C}_\infty)$-almost surely,
where and $\widehat\chi_\star$ is defined by \eqref{chitildedef}
with $\widetilde H$ everywhere replaced by~$\widehat H$. However, 
we have not made any serious attempt to carry out the details.

Surprisingly, unlike in the cases discussed in
Proposition~\ref{power-law}, $\widehat H$ takes a {\it unique\/}
functional form:
\begin{equation}
\label{ylogy}
\widehat H(y)=\sigma y\log y,
\end{equation}
where $\sigma>0$ is a parameter. This fact is established by
arguments similar to those used in the proof of
Proposition~\ref{power-law}. (As a by-product, we also get that
$t\mapsto\vartheta_t$ is slowly varying as $t\to\infty$.) An
interesting consequence of this is that, unlike in $\gamma<1$
situations, the variational problems for $\widehat\chi$ and
$\widehat\chi_\star$ factorize to one-dimensional problems (see
G\"artner and den Hollander \cite{GH99}).

\subsection{An application: Self-attractive random walks\label{attr}}
One of our original sources of motivation for this work have been
self-attractive path measures as models for ``squeezed polymers.''
Consider a polymer $S=(S_0,\dots,S_n)$ of length $n$ modeled by a
path of simple random walk with weight $\exp[\beta\sum_x
V(\ell_n(x))]$. Here $V\colon\Z\to (-\infty,0]$, and
$\ell_n(x)=\#\{k\leq n\colon S_k=x\}$ is the local time at~$x$.
Assuming that $V$ is convex and $V(0)=0$, e.g.,
$V(\ell)=-\ell^{\gamma}$ with $\gamma\in[0,1)$, the interaction
has an attractive effect. A large class of such functions $V$
(i.e., the completely monotonous ones) are the cumulant generating
functions of probability distributions on $[-\infty,0]$, like $H$
in~\eqref{Hdef}. Via the Feynman-Kac representation, this makes
the study of the above path measure essentially equivalent to the
study of the moments of a parabolic Anderson model. In fact, the
only difference is that for polymer models the time of the walk is
discrete.

We have no doubt that Theorem~\ref{momasy} extends to the
discrete-time setting. Hence, the endpoint $S_n$ of the polymer
should fluctuate on the scale $\alpha_n$ as in our Scaling
Assumption, which is $\alpha_n=n^\nu$ in the
$V(\ell)=-\ell^\gamma$ case. Since $\gamma\mapsto\nu$ is
decreasing, we are confronted with the counterintuitive fact that
the squeezing effect is the more extreme the ``closer'' is $V$ to
the linear function. This is even more surprising if one recalls
that for the boundary case $\gamma=1$, the Hamiltonian $\sum_x
V(\ell_n(x))$ is deterministic, and therefore the endpoint runs on
scale $n^{1/2}$. Note that, on the other hand, for $\gamma>1$,
which is the self-repellent case, it is known in $d=1$ (and
expected in dimensions $d=2$ and~$3$) that the scale of the
endpoint is a power larger than $1/2$. Hence, at least in low 
dimensions, there is an intriguing phase transition for the path 
scale at $\gamma=1$.

As a nice side-remark, the following model of an {\sl annealed
randomly-charged polymer} also falls into the class of models
considered above.  Consider an $n$-step simple random walk
$S=(S_0,\dots,S_n)$ with weight $e^{-\beta{\mathcal I}_n(S)}$,
where $\beta>0$ and
\begin{equation}
{\mathcal I}_n(S)=\sum_{0\le i<j\le n} \omega_i\omega_j\1\{S_i=S_j\}.
\end{equation}
Here $\omega=(\omega_i)_{i\in\N_0}$ is an i.i.d.\ sequence with a
symmetric distribution on ${\mathbb R}$ having variance one. Think
of $\omega_i$ as an electric charge at site $i$ of the polymer.
(For continuous variants of this model and more motivation see
e.g.\ Buffet and Pul\'e \cite{BP97}).

If the charges  equilibrate faster than the walk, the interaction
they effectively induce on the walk is given by the expectation
${E}(e^{-\beta{\mathcal I}_n(S)})$ and is thus of the above type
with
\begin{equation}
V(\ell)=-\log{E}\exp\bigl((\omega_0+\dots+\omega_\ell)^2\bigr),
\end{equation}
where ${E}$ denotes the expectation with respect to $\omega$. By
the invariance principle, we have $V(\ell)=-(1/2+o(1))\log\ell$ as
$\ell\to\infty$, which means that $V$ satisfies our Scaling
Assumption with $\alpha_n= (n/\log n)^{1/(d+2)}$. Hence, we can
identify the logarithmic asymptotics of the partition function
${\mathbb E}_0 \otimes E(e^{-\beta{\mathcal I}_n})$ and see that
the typical end-to-end distance of the annealed charged polymer
runs on the scale $\alpha_n$, i.e., the averaging over the charges
has a strong self-attractive effect.

\subsection{Relation to earlier work\label{relwork}}
General mathematical aspects of the problem \eqref{Anderson},
including the existence and uniqueness of solutions and a
criterion for intermittency [see \eqref{intermittency} and the
comments thereafter], were first addressed by G\"artner and
Molchanov \cite{GM90}. In a subsequent paper \cite{GM98}, the 
same authors  focused on the case of {\it double-exponential\/} 
distributions
\begin{equation}\label{doubleexp}
\prob(\xi(0)>x)\sim \exp\bigl\{-e^{x/\varrho}\bigr\},\qquad x\to\infty.
\end{equation}

For $0<\varrho<\infty$, the main contribution to $\langle
u(t,0)^p\rangle$ comes from islands in $\Z^d$ of asymptotically
finite size (which corresponds to a constant $\alpha_t$ in our
notation). When the upper tails of $\prob(\xi(0)\in\cdot)$ are yet
thicker (i.e., $\varrho=\infty$), e.g., when $\xi(0)$ is Gaussian,
then the overwhelming contribution to $\langle u(t,0)^p\rangle$
comes from very high peaks of $\xi$ concentrated at single sites.
(In a continuous setting the scaling can still be non-trivial, see
G\"artner and K\"onig \cite{GK98}, and G\"artner, K\"onig and
Molchanov \cite{GKM99}.) For thinner tails than double-exponential
(i.e., when $\varrho=0$, called the {\it almost bounded\/} case in
\cite{GM98}), the relevant islands grow unboundedly as
$t\to\infty$, i.e., $\alpha_t\to\infty$ in our notation. The
distribution \eqref{doubleexp} thus constitutes a certain critical
class for having a non-degenerate but still discrete spatial
structure.

The opposite extreme of tail behaviors was addressed by Donsker
and Varadhan \cite{DV79} (moment asymptotics) and by Antal
\cite{A95} (almost-sure asymptotics), see also \cite{A94}. The
distribution considered by these authors is $\xi(0)=0$ or
$-\infty$ with probability $p$ and $1-p$, respectively. The
analysis of the moments can be reduced to a self-interacting
polymer problem (see Subsection~\ref{attr}), which is essentially
the route taken by Donsker and Varadhan. In the almost-sure case,
the problem is a discrete analogue of the Brownian motion in a
Poissonian potential analyzed extensively by Sznitman in the
1990's using his celebrated method of enlargement of obstacles
(MEO), see Sznitman \cite{S98}.

The MEO bears on the problem \eqref{Anderson} because of the 
special form of the $\xi$~distribution: Recall the interpretation 
of points $z$ with $\xi(z)=-\infty$ as ``hard traps'' where the 
simple random walk is strictly killed. If ${\mathcal 
O}=\{z\in\Z^d\colon \xi(z)=-\infty\}$ denotes the trap region and 
$T_{\mathcal O}=\inf\{t>0\colon X(t)\in {\mathcal O}\}$ the first 
entrance time, then
\begin{equation}
\label{survprob}
u(t,z)=\P_z(T_{\mathcal O}>t),
\end{equation}
i.e., $u(t,z)$ is the survival probability at time $t$ for a walk 
started at $z$. In his thesis \cite{A94}, Antal derives a 
discrete version of the MEO and demonstrates its value in 
\cite{A94} and \cite{A95} by proving results which are (slight 
refinements of) our Theorems~\ref{momasy} and \ref{asasy} for 
$\gamma=0$ and $\alpha_t=t^{1/(d+2)}$.

The primary goal of this paper was to fill in the gap between the 
two regimes considered in \cite{GM98} and \cite{DV79} resp.\ 
\cite{A95}, i.e., we wanted to study the general case in which the 
diameter $\alpha_t$ of the relevant islands grows to infinity. We 
succeeded in doing that under the restrictions that the field is 
bounded from above and $\alpha_t$ diverges at least like a power 
of $t$. As already noted in Subsection~\ref{gamma=1}, in the 
boundary case $\alpha_t=t^{o(1)}$ (i.e., \hbox{$\gamma=1$}) 
another phenomenon occurs which cannot be handled in a unified 
manner; see the discussion of ``almost-bounded'' cases in the next 
subsection.

The technique of our proofs draws heavily on that of G\"artner and
K\"onig \cite{GK98} and G\"artner, K\"onig and Molchanov
\cite{GKM99}, however, non-trivial adaptations had to be made. An
interesting feature of this technique is the handle of the
compactification argument: We do not use folding (as Donsker and
Varadhan did in their seminal papers \cite{DV75} and \cite{DV79})
nor do we coarse-grain the field as is done in the MEO; instead,
we develop comparison arguments for Dirichlet eigenvalues in large
and small boxes. The task is in many places facilitated by
switching between the dual languages of Dirichlet eigenvalues {\sl
vs} local times of the simple random walk.

After this paper had been submitted, we learned that F.~Merkl and
M.~W\"uthrich had independently used rather similar techniques to
describe the scaling of the principal eigenvalue of the continuous
Dirichlet operator $-\Delta+(\log t)^{-2/d}V_\omega$ in
$[-t,t]^d$, where $V_\omega$ is the potential generated by
convoluting a shape function with the Poissonian cloud. (The
scaling of $V_\omega$ is chosen such that the eigenvalue is not
dominated solely by the potential, as in a certain sense happens
in the ``obstacle case.'') The first part of the results appeared
in Merkl and W\"uthrich~\cite{MW00}.

\subsection{Discussion and open problems\label{onourmethod}}
\indent (1) {\it ``Almost-bounded'' cases.\/\/}  As discussed in
Subsection~\ref{gamma=1}, the $\gamma=1$ case requires analyzing a
lower-order scale than considered in this paper. Interestingly,
the variational problem driving this scale coincides with that of
$\rho=0$ limit of the double exponential case; see
\eqref{doubleexp} and, e.g., G\"artner and den Hollander
\cite{GH99}. This makes us believe that the $\gamma=1$ case
actually reflects the {\it whole\/} regime of ``almost bounded''
but unbounded potentials, i.e., those interpolating between our
cases $\gamma<1$ and the double exponential distribution. (In all
these cases, we expect the following strategy of proof to be
universally applicable: identify the maximum of $\xi$ in a box of
size $t$ and, subtracting this term away, map the problem to the
effectively bounded case; see Subsection~\ref{gamma=1} for an
example.) For these reasons, we leave its investigation to future
work. \vspace{0.5mm}

(2) {\sl Generalized MEO.\/\/\/} Despite the fact that our current
technique circumvents the use of the MEO, it would be interesting
to develop its extension including other fields in our class (in
particular, those with $\gamma\not=0$). The main reason is that
this should allow for going beyond the leading order term.
However, the so called ``confinement property,'' which is the main
result of the MEO we cannot obtain, would require rather detailed
knowledge of the {\it shape\/} of the field that brings the main
contribution to the moments of $u(t,0)$ resp.\ to $u(t,0)$ itself.
Thus, while the MEO can help in controlling the ``probability
part'' of the statements \eqref{chi} and \eqref{asasyformula}, an
analysis of the minimizers in \eqref{chidef} and
\eqref{chitildedef} is also needed. The latter is expected to be
delicate in higher dimensions (in $d=1$ this task has fully been
carried out in Biskup and K\"onig \cite{BK98}). \vspace{0.5mm}

(3) {\sl Adding a drift.\/\/} An interesting open problem arises 
if a homogeneous drift term $\boldsymbol h\cdot\nabla u$ is added 
on the right-hand side of \eqref{Anderson}. This problem is 
considered hard (especially in $d\ge2$), since the associated 
Anderson Hamiltonian lacks self-adjoinedness with respect to the 
canonical inner product on $\ell^2(\Z^d)$. Self-adjointness can 
be restored if the inner product is appropriately modified; 
however, this case seems to be much more difficult to handle. One 
expects an interesting phase transition of the decay rate as 
$|\boldsymbol h|$ increases, but the rigorous understanding is 
rather poor at the moment. \vspace{0.5mm}

(4) {\sl Intermittency.\/\/} Our results imply intermittency for 
our model in the sense of asymptotic properties of positive 
moments of $u(t,0)$; see Remarks~\ref{intermitt1} and 
\ref{intermitt2}. The picture would round up very nicely if one 
could identify precisely the set of ``islands'' (or rather peaks) 
in the ``$\xi$~landscape,'' where the main contribution to 
$\langle u(t,0)\rangle$ resp.\ $u(t,0)$ comes from. At the 
moment, work of G\"artner, K\"onig and Molchanov \cite{GKM01} for 
the double-exponential distributions of the potentials is going 
on in this direction. Some additional complications stemming from 
$\alpha_t\to\infty$ can be expected in our present 
cases.\vspace{0.5mm}

(5) {\sl Correlation structure.\/\/} Another open problem concerns
the asymptotic correlation structure of the random field
$u(t,\cdot)$, as has been analyzed by G\"artner and den Hollander
\cite{GH99} in the case of the double-exponential distribution.
Also for answering this question, quite some control of the
minimizers in \eqref{chidef} and \eqref{chitildedef} is required.
Unfortunately, the compactification technique of \cite{GH99}
cannot be applied without additional work, since it seems to rely
on the discreteness of the underlying space in several important
places. As already alluded to, extension of this technique to
continuous space may also be relevant for the analysis of
\eqref{Anderson} with ``almost-bounded'' fields. \vspace{0.5mm}

\section{Preliminaries}
\label{prepare}
\vspace{2mm}\noindent 
In this section we first introduce some necessary
notation needed in the proof of Theorems~\ref{momasy}
and~\ref{asasy} and then prove Propositions~\ref{power-law}
and~\ref{tildechiident}. In the last subsection, we prove a claim
on the convergence of certain approximants to the variational
problem \eqref{chidef}.

\subsection{Feynman-Kac formula and Dirichlet eigenvalues\label{FKform}}
Our analysis is based on the link between the random-walk and
random-field descriptions provided by the Feynman-Kac formula. Let
$(X(s))_{s\in[0,\infty)}$ be the continuous-time simple random
walk on $\Z^d$ with generator $\kappa\DeltaD$. By $\P_z$ and
$\E_z$ we denote the probability measure resp.\ the expectation
with respect to the walk starting at $X(0)=z\in\Z^d$.

\subsubsection{General initial problem}
For any potential $V\colon \Z^d\to[-\infty,0]$, we denote by $u^V$
the unique solution to the initial~problem
\begin{equation}\label{AndersonV}
\begin{array}{rcll}
\displaystyle
\partial_t u(t,z)\!\!\! &=&\!\!\!\kappa \DeltaD u(t,z)+V(z)
u(t,z),\qquad
&(t,z)\in(0,\infty)\times \Z^d,\\
u(0,z)\!\!\!&=&\!\!\!1,&z\in\Z^d.
\end{array}
\end{equation}
Note that we have to set $u(t,z)\equiv0$ whenever $V(z)=-\infty$,
in order that \eqref{AndersonV} is well defined. The Feynman-Kac
formula allows us to express $u^V$ as
\begin{equation}\label{FK}
u^V(t,z)=\E_z\left[\exp\int_0^t
V\bigl(X(s)\bigr)\,ds\right],\qquad z\in\Z^d,\, t>0.
\end{equation}
Introduce the local times of the walk
\begin{equation}\label{localtimes}
\ell_t(z)=\int_0^t \1\{X(s)=z\}\,ds,\qquad z\in\Z^d,\, t>0,
\end{equation}
i.e., $\ell_t(z)$ is the amount of time the random walk has spent
at  $z\in\Z^d$ by time $t$. Note that $\int_0^t
V(X(s))\,ds=(V,\ell_t)$, where $(\cdot,\cdot)$ stands for the
inner product on $\ell^2(\Z^d)$.

In the view of \eqref{heu1}, of particular importance will be the
finite-volume version of \eqref{AndersonV} with Dirichlet boundary
condition. Let $R>0$ and let
$Q_R=[-R,R]^d\cap \Z^d $
be a box in $\Z^d$. The
solution of the initial-boundary value problem
\begin{equation}\label{initialboundary}
\begin{array}{rcll}
\displaystyle
\partial_t u(t,z)\!\!\! &=&\!\!\!\kappa \DeltaD u(t,z)+V(z)
u(t,z),\qquad &(t,z)\in(0,\infty)\times Q_R,\\
u(0,z)\!\!\!&=&\!\!\!1,&z\in Q_R,\\
u(t,z)\!\!\!&=&\!\!\!0,&t>0,\,z\notin Q_R,
\end{array}
\end{equation}
will be denoted by $u^V_R\colon [0,\infty)\times \Z^d\to[0,\infty)$.
Similarly to \eqref{FK}, we have the representation
\begin{equation}\label{FKDir}
u^V_R(t,z)=\E_z\Bigl[\exp\Bigl\{\int_0^t
V\bigl(X(s)\bigr)\,ds\Bigr\}\1\{\tau_R> t\}\Bigr], \qquad
z\in\Z^d,\, t>0,
\end{equation}
where $\tau_R$ is the first exit time from the set $Q_R$, i.e.,
\begin{equation}\label{taudef}
\tau_R=\inf\bigl\{t>0\colon X(t)\notin Q_R\bigr\}.
\end{equation}
Alternatively,
\begin{equation}
\label{altformula}
u^V_R(t,z)=\E_z\Bigl[e^{(V,\ell_t)}\1\bigl\{\supp(\ell_t)\subset
Q_R\bigr\}\Bigr],
\end{equation}
where we recalled \eqref{localtimes}. Note that,
for $0<r<R<\infty$,
\begin{equation}\label{umonoton}
u_r^V\leq u_R^V\leq u^V\qquad\mbox{in }[0,\infty)\times\Z^d,
\end{equation}
as follows by \eqref{FKDir} because $\{\tau_r> t\}\subset\{\tau_R>t\}$.

Apart from $u^V\!\!$, we also need the fundamental solution
$p_R^V(t,\cdot,z)$ of \eqref{initialboundary}, i.e., the solution
to \eqref{initialboundary} with $p_R^V(0,\cdot,z)=\delta_z(\cdot)$
instead of the second line. The Feynman-Kac representation is
\begin{equation}\label{FKp}
p_R^V(t,y,z)=\E_y \Bigl[e^{(V,\ell_t)}\1\bigl\{\supp (\ell_t)\subset
Q_R\bigr\}\1\bigl\{X(t)=z\bigr\}\Bigr]\qquad y,z\in\Z^d.
\end{equation}
Note that $\sum_{z\in Q_R}p_R^V(t,y,z)=u^V_R(t,y)$.

\subsubsection{Eigenvalue representations}
The second crucial tool for our proofs will be the principal (i.e.,
the largest) eigenvalue $\lambda_R^{\rm d}(V)$ of the operator
$\kappa\Delta^{\rm d}+V$ in $Q_R$ with Dirichlet boundary
condition. The Rayleigh-Ritz formula reads
\begin{equation}\label{discreteeigenv}
\lambdaD_R(V)=\sup\bigl\{(V,g^2)-\kappa \Vert\nablaD
g\Vert_{2}^2\colon g\in\ell^2(\Z^d), \|g\|_{2}=1,\supp(g)\subset
Q_R\bigr\}.
\end{equation}
Here $\nablaD$ denotes the discrete gradient.

Let $\lambda_1>\lambda_2\geq \lambda_3\geq \dots\geq\lambda_n$,
$n=\#Q_R$, be the eigenvalues of the operator $\kappa\DeltaD+V$ in
$\ell^2(Q_R)$ with Dirichlet boundary condition (some of them can
be $-\infty$). We also write $\lambda^{{\rm d},k}_R(V)=\lambda_k$
for the $k$-th eigenvalue to emphasize its dependence on the
potential and the box $Q_R$. Let $({\rm e}_k)_{k}$ be an
orthonormal basis in $\ell^2(Q_R)$ consisting of the corresponding
eigenfunctions ${\rm e}_k={\rm e}^{{\rm d},k}_{R}(V)$.
(Conventionally, ${\rm e}_k$ vanishes outside $Q_R$.) Then we have
the Fourier expansions
\begin{equation}\label{Fourierp}
p_R^V(t,y,z)=\sum_{k} e^{t\lambda_k} {\rm e}_k(y){\rm e}_k(z)
\end{equation}
and, by summing this over all $y\in Q_R$,
\begin{equation}\label{Fourieru}
u_R^V(t,\cdot)=\sum_{k}e^{t\lambda_k}({\rm e}_k,\1)_R \,{\rm e}_k(\cdot),
\end{equation}
where we used $(\cdot,\cdot)_R$ to denote the inner product in
$\ell^2(Q_R)$. Here and henceforth ``$\1$'' is the function taking
everywhere value $1$.

\subsection{Power-law scaling\label{sub-powlaw}}
\vspace{-6mm}
\begin{proofsect}{Proof of Proposition~\ref{power-law}}
Let $\widetilde H_t$ be the function given by
\begin{equation}
\label{H_t}
\widetilde H_t(\bdot)=\frac{\alpha_t^{d+2}}t H\biggl(\frac{t}{\alpha_t^d}
\,\bdot\biggr).
\end{equation}
By our Scaling Assumption, $\lim_{t\to\infty}\widetilde
H_t=\widetilde H$ on $[0,\infty)$. Note that both $\widetilde H_t$
and $\widetilde H$ are convex, non-positive and not identically
vanishing with value 0 at zero. Consequently, $\widetilde H_t$ and
$\widetilde H$ are continuous and strictly negative in
$(0,\infty)$. Moreover, by applying Jensen's inequality to the
definition of $H$, we have that $y\mapsto \widetilde H_t(y)/y$ and
$y\mapsto \widetilde H(y)/y$ are both non-decreasing functions.

Next we shall show that ${\alpha_{pt}}/{\alpha_{t}}$ tends to a
finite non-zero limit for all $p$. Let us pick a $y>0$ and a
$p\in(0,\infty)$ and consider the identity
\begin{equation}
\label{eqforH} \widetilde H_t\biggl(p\Bigl(
\frac{\alpha_t}{\alpha_{pt}}\Bigr)^d y\biggr)= p\Bigl(
\frac{\alpha_t}{\alpha_{pt}}\Bigr)^{d+2}\widetilde H_{pt}(y),
\end{equation}
which results by comparing \eqref{H_t} with the ``time'' parameter
interpreted once as $t$ and next time as $pt$. Invoking the monotonicity of
$y\mapsto\widetilde H_t(y)/y$, it follows that
\begin{equation}
p\Bigl(
\frac{\alpha_t}{\alpha_{pt}}\Bigr)^2 \widetilde H_{pt}(y)\ge
\widetilde H_t(py)\quad\text{whenever}\quad \alpha_t\ge\alpha_{pt}.
\end{equation}
This implies that $\alpha_{pt}/\alpha_{t}$ is bounded away from zero,
because we have
\begin{equation}
\label{bdonalpha}
\liminf_{t\to\infty}\Bigl(
\frac{\alpha_{pt}}{\alpha_t}\Bigr)^2
\ge \frac{p\widetilde H(y)}{\widetilde H(py)}\wedge 1>0,
\end{equation}
where ``$\wedge$'' stands for minimum. Since $p\in (0,\infty)$ was
arbitrary, $\alpha_{pt}/\alpha_{t}$ is also uniformly bounded, by
replacing $t$ with $t/p$.

Let $\phi(p)$ be defined for each $p$ as a subsequential limit of
${\alpha_{pt}}/{\alpha_{t}}$, i.e., $\phi(p)=\lim_{n\to\infty}
\alpha_{pt_n}/\alpha_{t_n}$ with some ($p$-dependent)
$t_n\to\infty$. By our previous reasoning $\phi(p)^{-1}$ is
non-zero, finite and, for all $y>0$, it solves for $z$ in the
equation
\begin{equation}
\label{power-laweq}
\widetilde H\bigl(p z^d y\bigr)=p z^{d+2}\widetilde H(y).
\end{equation}
Here we were allowed to pass to the limiting function $\widetilde
H$ on the left-hand side of \eqref{eqforH} because $\widetilde H$
is continuous and the scaling limit \eqref{Hscaling} is uniform on
compact sets in $(0,\infty)$. But $z\mapsto\widetilde H(p z^d
y)/z^d$ is non-decreasing while $z\mapsto pz^2\widetilde H(y)$ is
{\it strictly\/} decreasing, so the solution to
\eqref{power-laweq} is unique. Hence, the limit
$\phi(p)=\lim_{t\to\infty}\alpha_{pt}/\alpha_{t}$ exists in
$(0,\infty)$ for all $p\in(0,\infty)$.

It is easily seen that $\phi$ is multiplicative on $(0,\infty)$,
i.e., $\phi(pq)=\phi(p)\phi(q)$. Since $\phi(p)\ge1$ for $p\ge1$,
by the same token we also have that $p\mapsto\phi(p)$ is
non-decreasing. These two properties imply that
$\phi(2^n)=\phi(2)^n$ and that $\phi(2)^{\frac nm} \le\phi(p)\le
\phi(2)^{\frac {n+1}m}$ for any $p>0$, and $m$, $n$ integer such
that $2^n\le p^m< 2^{n+1}$. Consequently, $\phi(p)=p^{\nu}$ with
$\nu=\log_2 \phi(2)$. By plugging this back into
\eqref{power-laweq} and setting $y=1$ we get that
\begin{equation}
\widetilde H\bigl(p^{1-d\nu}\bigr)=\widetilde H(1)\, p^{1-(d+2)\nu}.
\end{equation}
The claims \eqref{perfscalforH} and \eqref{scalingexp1} are thus
established by putting $\gamma(1-d\nu)=1-(d+2)\nu$, which is
\eqref{nudef}. Clearly, $\gamma\in[0,1]$, in order to have the
correct monotonicity properties of $y\mapsto\widetilde H(y)$ and
$y\mapsto\widetilde H(y)/y$.

To prove also the second statement in \eqref{scalingexp1}, we first write
\begin{equation}
\alpha_{2^N}=\alpha_1\prod_{m=0}^{N-1}
\frac{\alpha_{2^{m+1}}}
{\alpha_{2^m}}
\end{equation}
which, after taking the logarithm, dividing by $\log 2^N$, and noting that
${\alpha_{2^{m+1}}}/{\alpha_{2^m}}\to \phi(2)$ as $m\to\infty$,
allows us to conclude that
\begin{equation}
\lim_{N\to\infty}
\frac{\log\alpha_{2^N}}{\log 2^N}=\log_2\phi(2)=\nu.
\end{equation}
The limit for general $t$ is then proved again by sandwiching $t$
between $2^{N-1}$ and $2^N$ and invoking the monotonicity of
$t\mapsto\alpha_t$.
\end{proofsect}

\subsection{Relation between $\boldsymbol\chi$ and
$\boldsymbol{\widetilde\chi}$} \label{tildechiproof} \vspace{-5mm}
\begin{proofsect}{Proof of Proposition~\ref{tildechiident}}
Suppose $H$ is in the $\gamma$-class and define $\nu$ as in
Proposition~\ref{power-law}. Suppose $\chi\not=0,\infty$ (for a proof of
this statement, see Proposition~\ref{chiprop}). The argument hinges on 
particular
scaling properties of the functionals $\psi\mapsto{\mathcal
L_R}(\psi)$ and $\psi\mapsto\lambda_R(\psi)$, which enable us
to convert \eqref{chidef} into \eqref{chitildedef}. Given $\psi\in
C^-(R)$, let us for each $b\in(0,\infty)$ define $\psi_b\in
C^-(bR)$ by
\begin{equation}
\label{psitransf}
\psi_b(x)=\frac 1{b^2}\psi\left(\frac xb\right).
\end{equation}
Then we have
\begin{equation}
\label{Llscaling}
{\mathcal L_{bR}}(\psi_b)=b^{\frac1\nu-2}{\mathcal L_R}(\psi)
\qquad\text{and}\qquad \lambda_{bR}(\psi_b)=b^{-2}\lambda_R(\psi),
\end{equation}
where in the first relation we used that $\psi_b$ can be converted
into $\psi$ in \eqref{calLdef} by substituting
$b^{2/(1-\gamma)}f(\cdot/ b)$ in the place of $f(\,\cdot\,)$;
the second relation is a result of a simple spatial scaling of the
first line in \eqref{lambdadef}. Note that $\frac1\nu-2\ge1>0$.

Let $\psi^{(n)}\in C^-(R_n)$ be a minimizing sequence of the
variational problem in \eqref{chiident}. Suppose,
without loss of generality, that ${\mathcal
L_{R_n}}(\psi^{(n)})\to \bar{\mathcal L}$ and
$\lambda_{R_n}(\psi^{(n)})\to\bar\lambda$. Then we have
\begin{equation}
\label{1st}
\chi= \bar{\mathcal L}-\bar\lambda.
\end{equation}
Now pick any $b\in(0,\infty)$ and consider instead the sequence
$(\psi^{(n)}_b)$.
Clearly,
\begin{equation}
\chi\leq \lim_{n\to\infty} \left[{\mathcal
L_{bR_n}}(\psi^{(n)}_b)-\lambda_{bR_n}(\psi^{(n)}_b)\right]
=b^{\frac1\nu-2} \bar{\mathcal L}-b^{-2}\bar\lambda
\end{equation}
for all $b$. By \eqref{1st}, the derivative of the right-hand
side must vanish at $b=1$, i.e.,
\begin{equation}
\label{2nd}
\left(\textstyle\frac1\nu-2\right)\bar{\mathcal L}+2\bar\lambda=0.
\end{equation}
By putting \eqref{1st} and \eqref{2nd} together, we easily compute that
\begin{equation}
\label{Lchi}
\bar{\mathcal L}=2\nu \chi.
\end{equation}

Note that while $b\mapsto {\mathcal L}_{bR}(\psi_b)$ is strictly
increasing,  $b\mapsto\lambda_{bR}(\psi_b)$ is strictly
decreasing. This allows us to recast \eqref{chiident} as
\begin{equation}
\label{alter}
\chi=\bar{\mathcal
L}+\inf_{R>0}\inf\left\{-\lambda_R(\psi)\colon\psi\in
C^-(R),\,{\mathcal L}_R(\psi)\le\bar{\mathcal L}\right\}.
\end{equation}
Indeed, we begin by observing that ``$\leq$'' holds in
\eqref{alter}, as is verified by pulling $\bar{\mathcal L}$ inside
the bracket, replacing it with ${\mathcal L}_R(\psi)$, and
dropping the last condition. To prove the ``$\geq$'' part,
note that the above sequence $(\psi^{(n)}_b)$ for $b<1$ eventually
fulfills the last condition in \eqref{alter} because ${\mathcal
L}_{bR_n}(\psi^{(n)}_b)\to b^{\frac1\nu-2}\bar{\mathcal
L}<\bar{\mathcal L}$. Since $\lambda_{bR_n}(\psi^{(n)}_b)\to
b^{-2}\bar\lambda$,  the right-hand side of \eqref{alter} is
no more than $\bar{\mathcal L}-b^{-2}\bar\lambda$ for any $b<1$.
Taking $b\uparrow 1$ and recalling \eqref{1st} proves the equality
in \eqref{alter}.

With \eqref{alter} in the hand we can finally prove
\eqref{chichirel}. By using $\psi_b$ instead of $\psi$ in
\eqref{alter}, the condition ${\mathcal L}_R(\psi)\le\bar{\mathcal
L}$ becomes ${\mathcal L}_{R}(\psi)\le
b^{\frac1\nu-2}\bar{\mathcal L}$ and the factor  $b^{-2}$ appears
in front of the infimum. Thus, setting
$b^{\frac1\nu-2}\bar{\mathcal L}=d$, which by \eqref{Lchi}
requires that
\begin{equation}
b=\left(\frac {2\nu\chi}{d}\right)^{\frac{\nu}{1-2\nu}},
\end{equation}
(note that $b\not=0,\infty$) and invoking \eqref{Lchi}, we recover
the variational problem \eqref{chitildedef}. Therefore,
\begin{equation}
\chi=\bar{\mathcal L}+b^{-2}\widetilde\chi=2\nu\chi+\left(\frac
{2\nu\chi}{d}\right)^{-\frac{2\nu}{1-2\nu}} \widetilde\chi.
\end{equation}
From this, \eqref{chichirel} follows by simple algebraic
manipulations. The claim  $\widetilde\chi\in(0,\infty)$ is a
consequence of \eqref{chichirel} and the fact that
$\chi\in(0,\infty)$.
\end{proofsect}

\subsection{Approximate variational problems\label{AVP}}
The proof of Theorem~\ref{momasy} will require some technical 
approximation  properties of the variational problem 
\eqref{chidef}. These are stated in Proposition~\ref{chiprop} 
below. The reader may gain more motivation for digesting the 
proof by reading first Subsection~\ref{momlower}.

Let $\chi_R$ be the finite-volume counterpart of $\chi$:
\begin{equation}\label{chiRdef}
\chi_R=\inf\bigl\{\I(f)-{\mathcal H}_R(f)\colon f\in \F_R\bigr\},\qquad R>0.
\end{equation}
Suppose $H$ is in the $\gamma$-class and introduce
the following quantities: In the case $\gamma\in(0,1)$, let
\begin{equation}
\label{chistar}
\chi_{R}^\star(M)=\inf\bigl\{\I(f)-{\mathcal H}_R(f\wedge M)\colon f\in
\F_{R}\bigr\},\qquad M>0,
\end{equation}
for any $R>0$. For $\gamma=0$ and any $R>0$, let
\begin{equation}
\label{chicross}
\chi_R^\#(\eps)=\inf\bigl\{\I(f)-\widetilde H(1)|\{f>\eps\}|\colon f\in
\F_{R}\bigr\},\qquad 0<\eps\ll R.
\end{equation}
The needed relations between $\chi$, $\chi_R$, $\chi_{R}^\star(M)$
and $\chi_R^\#(\eps)$ are summarized as follows:

\begin{proposition}
\label{chiprop}
Let $H$ be in the $\gamma$-class and let $\chi$ be as in
\eqref{chidef}. Then

(1) $\chi\in(0,\infty)$.

(2) For $\gamma\in(0,1)$ and any $R>0$, $\lim_{M\to\infty}
\chi_{R}^\star(M)=\chi_R$.

(3) For $\gamma=0$ and any $R>0$, $\lim_{\eps\downarrow0}\chi_R^\#(
\eps)=\chi_R$.
\end{proposition}

\begin{proofsect}{Proof of (1) and (2)}
Assertion (1) for $\gamma=0$ is well-known. Assume that
$\gamma\in(0,1)$ and observe that, due to the perfect scaling
properties of both $f\mapsto\I(f)$ and $f\mapsto{\mathcal
H}_R(f)$, \eqref{chiRdef} can alternatively be written as
\begin{equation}
\chi_R=\inf\bigl\{ R^{-2}\I(f)-R^{d(1-\gamma)}{\mathcal H}_1(f)
\colon f\in \F_1\bigr\}.
\end{equation}
Let $(\lambda_1,\widehat g)$ be the principal eigenvalue resp.\ an
associated eigenvector of $-\Delta$ in $[-1,1]^d$ with
Dirichlet boundary condition. Then $\I(\widehat
g^2)=\kappa\lambda_1\not=0,\infty$, which means that
\begin{equation}
\label{barchiR}
\chi_R\le R^{-2}\kappa\lambda_1-R^{d(1-\gamma)}\widetilde H(1)
\int |\widehat g|^{2\gamma}=:\bar\chi_R.
\end{equation}
Since $\widehat g$ is continuous and bounded, the integral is finite,
whereby $\chi\le\inf_{R>0}\bar\chi_R<\infty$.

Claim (2) and the remainder of (1) are then simple consequences of the
following
observation, whose justification we defer to the end of this proof:
\begin{equation}
\label{infI} \inf\bigl\{\I(f)\colon f\in\F_R,\, \Vert f\1_{\{f\ge
M\}}\Vert_1\ge\eps\bigr\}\ge \kappa\, \frac\eps2\Bigl(\frac{
M}{8\pi_d}\Bigr)^{2/d}, \quad\,\,\, R,\eps>0, \,\,M\ge 8\pi_d
d^d/R^d,
\end{equation}
where $\pi_d$ is the volume of the unit sphere in $\R^d$. Indeed,
to get that $\chi$ is non-vanishing, set $\eps=1/2$ and choose $M$
such that the infimum in \eqref{infI} is strictly larger than
$-\widetilde H(1)M^{\gamma-1}/2$ for all $R\ge1$. Clearly, $M$ is
finite, so $C:=-\widetilde H(1)M^{\gamma-1}/2>0$. Then for any
$f\in\F_R$ either $\Vert f\1_{\{f\ge M\}}\Vert_1\ge 1/2$, which
implies $\I(f)\ge C$, or $\Vert f\1_{\{f\ge M\}}\Vert_1<1/2$ which
implies
\begin{equation}
-{\mathcal H}_R(f)\ge-\widetilde H(1)\int f^\gamma\, \1_{\{f<
M\}}\ge -\widetilde H(1)M^{\gamma-1}\int f\,\1_{\{f<
M\}}\ge-\widetilde H(1) M^{\gamma-1}/2=C.
\end{equation}
Thus, in both cases, $\I(f)-{\mathcal H}_R(f)\ge C>0$ independent
of $R$. Since $R\mapsto\chi_R$ is decreasing, the restriction to
$R\ge1$ is irrelevant which finishes part (1).

To prove also part (2), note first that $\chi_R^\star(M)\le\chi_R$
for all $M>0$. Given $\eps>0$, let $M\ge1$ be such that the infimum
in \eqref{infI} is larger than $\bar\chi_R$ in \eqref{barchiR}.
Consider \eqref{chistar} restricted to $f\in\F_R$ with
$\Vert f\1_{\{f\ge M\}}\Vert_1<\eps$. Since for any such $f$
\begin{multline}
-{\mathcal H}_R(f\wedge M)\ge-\widetilde H(1)\int f^\gamma\, \1_{\{f< M\}}
\ge -{\mathcal H}_R(f)+\widetilde H(1)\int f^\gamma\, \1_{\{f\ge M\}}
\\
\ge
-{\mathcal H}_R(f)+\widetilde H(1)\int f\, \1_{\{f\ge M\}}\ge
-{\mathcal H}_R(f)+\widetilde H(1)\eps,
\end{multline}
the restricted infimum is no less than $\chi_R+\widetilde
H(1)\eps$. Therefore,
$\chi_R^\star(M)\ge\bar\chi_R\wedge(\chi_R+\widetilde H(1)\eps)$,
which by $\eps\downarrow0$ and \eqref{barchiR} proves part (2) of
the claim.

It remains to prove \eqref{infI}. To that end,
denote the infimum by $\Psi_R(\eps,M)$ and note that
\begin{equation}
\label{PsiR}
\Psi_R(\eps,M)=R^{-2}\Psi_1(\eps,M R^d).
\end{equation}
Indeed, denoting $f^*(\,\cdot\,)=R^d f(\cdot R)$ for any
$f\in\F_R$, we have $f^*\in\F_1$, $\I(f^*)=R^2\I(f)$, and $\Vert
f^*\1_{\{f^*\ge MR^d\}}\Vert_1=\Vert f\1_{\{f\ge M\}}\Vert_1$,
whereby \eqref{PsiR} immediately follows. Since
$R^{-2}(MR^d)^{2/d}=M^{2/d}$, it suffices to prove \eqref{infI}
just for $R=1$.

Recall that the operator $-\Delta$ on $[-1,1]^d$ with Dirichlet
boundary condition has a compact resolvent, so its spectrum
$\sigma(-\Delta)$ is a discrete set of finitely-degenerate
eigenvalues. For each $k\in{\mathbb N}$, define the function
\begin{equation}
\varphi_k(x)=\begin{cases}
\cos\bigl(\frac\pi{2} kx\bigr)
\quad&\text{if }k\text{ is odd},\\
\sin\bigl(\frac\pi{2} kx\bigr) \quad&\text{if }k\text{ is even}.
\end{cases}
\end{equation}
Then $\sigma(-\Delta)=\{\pi^2|k|_2^2/4\colon k\in{\mathbb
N}^d\}$, with $|k|_2^2=k_1^2+\dots+k_d^2$ and the eigenvectors
given as $\omega_k=\varphi_{k_1}\otimes\dots\otimes\varphi_{k_d}$.
Note that the latter form a (Fourier) basis in $L^2([-1,1]^d)$.

Let $\eps>0$ and $M>0$ be fixed. Let $r$ be such that
$8\pi_dr^d=M$. Note that $r\ge d$. Pick a function $f\in\F_1$ such
that $ \Vert f\1_{\{f\ge M\}}\Vert_1\ge\eps$ and let $g=\sqrt f$.
Let $g_1$ resp.\ $g_2$ be the normalized projections of $g$ onto
the Hilbert spaces generated by $(\omega_k)$ with $|k|_2\le r$
resp.\ $|k|_2>r$. Then $g=a_1 g_1+a_2 g_2$ with
$|a_1|^2+|a_2|^2=1$. We claim that $\Vert g_1\Vert_\infty\le\sqrt
M/2$. Indeed, $g_1=\sum_k c_k\omega_k$ where
$(c_k)\in\ell^2({\mathbb N}^d)$ is such that $c_k=0$ for all
$k\in{\mathbb N}^d$ with $|k|_2>r$ and
\begin{equation}
\label{g1bound}
\Vert g_1 \Vert_\infty\le \sum_k
|c_k|\Vert\omega_k\Vert_\infty\le \sqrt{\#\{k\colon
c_k\not=0\}}\le \sqrt{2\pi_dr^d}=\sqrt M/2.
\end{equation}
Here we used that $\Vert\omega_k\Vert_\infty\le 1$, then we
applied Cauchy-Schwarz inequality and noted that $(c_k)$ is
normalized to one in $\ell^2({\mathbb N}^d)$, because $\Vert
\omega_k\Vert_2=1$ for all $k\in{\mathbb N}^d$. The third
inequality follows by the observation $\#\{k\colon c_k\not=0\}\le
\pi_d (r+1)^d/2d\le 2\pi_d r^d$ implied by $r\ge d$.

Let $x$ be such that $g(x)\ge\sqrt M$. Then we have
$\sqrt M\le g(x)\le |g_1(x)|+|a_2||g_2(x)|$.
Using \eqref{g1bound}, we derive that $|a_2||g_2(x)|\ge \sqrt M/2$,
whereby we have that $g(x)\le 2|a_2||g_2(x)|$. This gives us the bound
\begin{equation}
\label{one} \eps\le\Vert f\1_{\{f\ge M\}}\Vert_1=\Vert
g\1_{\{g\ge\sqrt M\}}\Vert_2^2\le 4|a_2|^2\Vert g_2
\Vert_2^2=4|a_2|^2,
\end{equation}
i.e., $|a_2|^2\ge\eps/4$. On the other hand,
\begin{equation}
\label{two}
\I(f)=\kappa\Vert\nabla g\Vert_2^2\ge \kappa|a_2|^2 \Vert\nabla
g_2\Vert_2^2\ge\kappa|a_2|^2\frac{\pi^2}4 r^2.
\end{equation}
where we used that $g_1\bot g_2$ and that $g_2$ has no overlap with
$\omega_k$ such that $|k|_2\le r$. By putting \eqref{one} and
\eqref{two} together and noting that $\pi^2/16\ge 1/2$,
\eqref{infI} for $R=1$ follows.
\end{proofsect}

\begin{proofsect}{Proof of (3)}
Let $\eps\ll(2R)^d$ and consider $f\in\F_R$. Let $g=\sqrt f$ and
define $g_\eps=(g-\sqrt\eps)\1\{g\ge\sqrt\eps\}$. By a
straightforward calculation, $\Vert g_\eps\Vert_2^2\ge
1-2\eps(2R)^d-2\sqrt{\eps(2R)^d}$. Let $f_\eps=(g_\eps/\Vert
g_\eps\Vert_2)^2$. Then $\I(f)\ge \Vert
g_\eps\Vert_2^2\,\I(f_\eps)$, while
$|\{f>\eps\}|=|\{f_\eps>0\}|$.
This implies that $\chi^\#_R(\eps)\ge\chi_R(1-O(\sqrt\eps))$.
Since $\chi_R^\#(\eps)\le\chi_R$, the proof is finished.
\end{proofsect}

\section{Proof of Theorems~\ref{momasy} and~\ref{Lifshitz}}
\label{proofmom}\noindent
We begin by deriving the logarithmic asymptotics for the moments of
$u(t,0)$ as stated in Theorem~\ref{momasy}. The proof is divided into two
parts: we separately prove the lower bound and the upper bound.
Whenever convenient, we write $\alpha(t)$ instead of $\alpha_t$.

\subsection{The lower bound\label{momlower}}
We translate the corresponding proof of \cite{GK98} into the
discrete setting. Let $u$ denote the solution to \eqref{Anderson},
denoted by $u^\xi$ in Section~\ref{prepare}. Similarly, let $u_R$
stand for $u_R^\xi$ for any $R>0$. Fix $p\in(0,\infty)$, $R>0$, and
consider the box $Q_{R\alpha(pt)}=
[-R\alpha(pt),R\alpha(pt)]^d\cap \Z^d$. Note that
$\#Q_{R\alpha(pt)}=e^{o(t\alpha_{pt}^{-2})}$ as $t\to\infty$.
Recall that $u_{R\alpha(pt)}(t,\cdot)=0$ outside
$Q_{R\alpha(pt)}$ and that $(\cdot,\cdot)$ denotes the inner
product in $\ell^2(\Z^d)$. Our first observation is the following.

\begin{lemma} As $t\to\infty$,
\begin{equation}
\label{start}
\bigl\langle u(t,0)^p\bigr\rangle\geq e^{o(t\alpha_{pt}^{-2})}
\bigl\langle (u_{R\alpha(pt)}(t,\cdot),\1)^p\bigr\rangle.
\end{equation}
\end{lemma}
\begin{proofsect}{Proof}
In the case $p\geq 1$, use the shift-invariance of $z\mapsto
u(t,z)$, Jensen's inequality, and the monotonicity assertion
\eqref{umonoton} to obtain
\begin{equation}
\label{startproof}
\begin{aligned}
\bigl\langle u(t,0)^p\bigr\rangle&=\Bigl\langle \frac
1{\#Q_{R\alpha(pt)}}\sum_{z\in
Q_{R\alpha(pt)}}
u(t,z)^p\Bigr\rangle\\
&\geq \Bigl\langle\Bigl(\frac 1{\#Q_{R\alpha(pt)}}\sum_{z\in
Q_{R\alpha(pt)}}
u(t,z)\Bigr)^p\Bigr\rangle\geq e^{o(t\alpha_{pt}^{-2})}
\bigl\langle (u_{R\alpha(pt)}(t,\cdot),\1)^p\bigr\rangle.
\end{aligned}
\end{equation}
In the case $p<1$, instead of Jensen's inequality we apply
\begin{equation}
\sum_{i=1}^n x_i^p\geq \Big(\sum_{i=1}^n x_i\Big)^p,
\qquad x_1,\dots,x_n\geq 0,\, n\in\N,
\end{equation}
to deduce similarly as in (\ref{startproof}) that
\begin{equation}
\begin{aligned}
\bigl\langle u(t,0)^p\bigr\rangle &=e^{o(t\alpha_{pt}^{-2})}
\Big\langle \sum_{z\in Q_{R\alpha(pt)}} u(t,z)^p\Big\rangle\\
&\geq e^{o(t\alpha_{pt}^{-2})}
 \Big\langle\Big(\sum_{z\in
Q_{R\alpha(pt)}}u(t,z)\Big)^p\Big\rangle\geq e^{o(t\alpha_{pt}^{-2})}
\big\langle (u_{R\alpha(pt)}(t,\cdot),\1)^p\big\rangle.
\end{aligned}
\end{equation}
\end{proofsect}

The following Lemma~\ref{largedeviat} carries out the necessary
large-deviation arguments for the case $p=1$.
Lemma~\ref{preduction} then reduces the proof of arbitrary $p$ to
the case $p=1$. Recall the ``finite-$R$'' version $\chi_R$ of
\eqref{chidef} defined in \eqref{chiRdef}.

\begin{lemma}\label{largedeviat}
Let $R>0$. Then for $t\to\infty$,
\begin{equation}\label{ldpu}
-\chi_R+o(1)\leq\frac{\alpha_t^2}t\log\left\langle
(u_{R\alpha(t)}(t,\cdot),\1) \right\rangle\leq-\chi_{3R}+o(1),
\end{equation}
\begin{equation}\label{ldplambda}
\frac{\alpha_t^2}t\log\Bigl\langle
\sum_{k}e^{t\lambda_{R\alpha(t)}^{{\rm d},k}(\xi)}\Bigr\rangle
\leq-\chi_{3R} +o(1).
\end{equation}
\end{lemma}

\begin{lemma}\label{preduction}
Let $R>0$. Then for $t\to\infty$,
\begin{equation}\label{pred}
\left\langle (u_{R\alpha(pt)}(t,\cdot),\1)^p\right\rangle\geq
e^{o(t\alpha_{pt}^{-2})} \left\langle
(u_{R\alpha(pt)}(pt,\cdot),\1)\right\rangle.
\end{equation}
\end{lemma}
\vspace{0.1cm}

Lemmas~\ref{start}, \ref{largedeviat}, and~\ref{preduction} make
the proof of the lower bound immediate:

\begin{proofsect}{Proof of Theorem~\ref{momasy}, lower bound}
By combining  \eqref{start}, \eqref{pred} and the left inequality
in \eqref{ldpu} for $pt$ instead of $t$, we see that
$(\alpha_{pt}^2/pt) \log \langle u(t,0)^p\rangle\ge-\chi_R+o(1)$.
Since $\lim_{R\to\infty}\chi_R=\chi$, the left-hand side of
\eqref{chi}, with ``$\liminf$'' instead of ``$\lim$,'' is bounded
below by $-\chi$. By Proposition~\ref{chiprop}(1), $\chi$
positive, finite and non-zero.
\end{proofsect}

The remainder of this subsection is devoted to the proof of the
two lemmas.

\begin{proofsect}{Proof of Lemma~\ref{largedeviat}}
Recall the notation of Subsection~\ref{FKform}. By taking the
expectation over $\xi$ (and using that $\xi$ is an i.i.d.\ field)
and recalling \eqref{altformula}, we have for any $z\in
Q_{R\alpha(t)}$ that
\begin{multline}
\label{lowbound}
\qquad\bigl\langle u_{R\alpha(t)}(t,z)\bigr\rangle=
\Bigl\langle\E_z \bigl[e^{(\xi,\ell_t)}
\1\{\tau_{R\alpha(t)}>t\}\bigr]\Bigr\rangle=\E_z\Bigl[\prod_{y\in\Z^d}
\bigl\langle e^{\ell_t(y)\xi(y)}\bigr\rangle\1\{\tau_{R\alpha(t)}>
t\}\Bigr]\\
=\E_z
\Bigl[\exp\Bigl\{\sum_{y\in\Z^d}H\bigl(\ell_t(y)\bigr)
\Bigr\}\1\bigl\{\supp(\ell_t)
\subset Q_{R\alpha(t)}\bigr\}\Bigr],\quad
\end{multline}
Consider
the scaled version
$\bar\ell_t\colon\R^d\to[0,\infty)$ of the local times
\begin{equation}
\quad
\bar\ell_t(x)=\frac{\alpha_t^d}t
\ell_t\bigl(\lfloor x\alpha_t\rfloor\bigr),\qquad
x\in\R^d.
\end{equation}
Let $\widetilde\F$ be the space of all non-negative Lebesgue
almost everywhere continuous functions in $L^1(\R^d)$ with a
bounded support. Clearly, $\F\subset\widetilde\F$ and
$\bar\ell_t\in\widetilde\F$. Introduce the functional
$\skrih^{(t)}\colon\widetilde\F\to [-\infty,0]$, assigning each
$f\in\widetilde\F$ the value
\begin{equation}
\label{curledHt}
\skrih^{(t)}(f)=\int_{\R^d}\widetilde H_t\bigl(f(x)\bigr)\,dx,
\end{equation}
where we recalled \eqref{H_t}. Substituting $\bar\ell_t$ and
$\skrih^{(t)}$ into \eqref{lowbound}, we obtain
\begin{equation}
\bigl\langle (u_{R\alpha(t)}(t,\cdot),\1)\bigr\rangle= \sum_{z\in
Q_{R\alpha(t)}}\E_z \Bigl[\exp\Bigl\{\frac
t{\alpha_t^2}\skrih^{(t)}\left(\bar\ell_t\right)
\Bigr\}\1\bigl\{\supp(\bar\ell_t) \subset [-R,R+\alpha_t^{-1}]^d
\bigr\}\Bigr].
\end{equation}
Using shift-invariance and the fact that $\skrih^{(t)}(f)\le
\skrih^{(t)}(f\wedge M)$ for any $M>0$, we have
\begin{multline}
\label{ldpappl}
\E_0\Bigl[\exp\Bigl\{\frac t{\alpha_t^2}\skrih^{(t)}
\left(\bar\ell_t\right) \Bigr\}\1\bigl\{\supp(\bar\ell_t) \subset
[-R,R]^d\bigr\}\1\{\bar\ell_t\le M\} \Bigr]\leq\bigl\langle
(u_{R\alpha(t)}(t,\cdot),\1)\bigr\rangle\\ \leq
e^{o(t\alpha_t^{-2})}\, \E_0\Bigl[\exp\Bigl\{\frac
t{\alpha_t^2}\skrih^{(t)} \left(\bar\ell_t\wedge M\right)
\Bigr\}\1\bigl\{\supp(\bar\ell_t) \subset [-3R,3R]^d\bigr\}\Bigr].
\end{multline}

It is well known that the family of scaled local times
$(\bar\ell_t)_{t>0}$ satisfies a weak large-deviation principle on
$L^1(\R^d)$ with rate $t\alpha_t^{-2}$ and rate function $\I$
defined in \eqref{Idef}. This fact has been first derived by
Donsker and Varadhan \cite{DV79} for the discrete-time random
walk; for the changes of the proof in the continuous time case we
refer to Chapter~4 of the monograph by Deuschel and
Stroock~\cite{DS89}. The large-deviation principle allows us to
use Varadhan's integral lemma to convert both bounds in
\eqref{ldpappl} into corresponding variational formulas. Note
that, if both $\I$ and ${\mathcal H}$ are appropriately extended
to $L^1([-R,R]^d)$, all infima \eqref{chiRdef}, \eqref{chistar}
and \eqref{chicross} can be taken over $f\in L^1([-R,R]^d)$ with
the same result. In the sequel, we have to make a distinction
between the cases $\gamma\in(0,1)$ and $\gamma=0$.

In the case $\gamma\in(0,1)$, our Scaling Assumption implies that,
for every $M>0$, $f\mapsto\skrih(f)$ is continuous and
$\skrih^{(t)}$ converges to $\skrih$ uniformly on the space of all
measurable functions $[-R,R]^d\to [0,M]$ with $L^\infty$ topology.
Indeed, for any such function $f$ and any $\eps>0$, the integral
\eqref{curledHt} can be split into
$\skrih^{(t)}(f\1_{\{f>\eps\}})$ and
$\skrih^{(t)}(f\1_{\{0<f\le\eps\}})$. The former then converges
uniformly to $\skrih(f\1_{\{f>\eps\}})$, while the latter can be
bounded as
\begin{equation}
\label{auxineq}
0\ge \skrih^{(t)}\bigl(f\1_{\{0<f\le \eps\}}\bigr)
\ge \widetilde H_t(\eps)\bigl|\{0<f\le\eps\}\bigr| \ge (2R)^d
\widetilde H_t(\eps),
\end{equation}
where we invoked the monotonicity of $y\mapsto\widetilde H_t(y)$.
Taking $\eps\downarrow0$ proves that this part is negligible for
$\skrih^{(t)}(f)$ and, if $t\to\infty$ is invoked before
$\eps\downarrow0$, it also shows that
$\skrih(f\1_{\{f>\eps\}})\to\skrih(f)$ uniformly in $f$ as
$\eps\downarrow0$. Having verified continuity, Varadhan's lemma
(and $M\to\infty$) readily outputs the left inequality in
\eqref{ldpu}, while on the right-hand side it yields a bound in
terms of the quantity $\chi_{3R}^\star(M)$ defined in
\eqref{chistar}. By Proposition~\ref{chiprop}(2),
$\chi_{3R}^\star(M)$ tends to $\chi_{3R}$ as $M\to\infty$, which
proves the inequality on the right of \eqref{ldpu}.

In the case $\gamma=0$, the lower bound goes along the same line,
but we have to be more careful with \eqref{auxineq}, since
$\lim_{\eps\downarrow0}\lim_{t\to\infty}\widetilde H_t(\eps)\not=0$
in this case. Let us estimate
\begin{multline}
\skrih^{(t)}(f)=\skrih^{(t)}\bigl(f\1_{\{0<f\le\eps\}}\bigr)
+\skrih^{(t)}\bigl(f\1_{\{f>\eps\}}\bigr)\ge \widetilde
H_t(\eps)\bigl|\{0<f\le\eps\}\bigr|+\skrih^{(t)}\bigl(f\1_{\{f>\eps\}}\bigr)\\
\ge\skrih(f)-\bigl|\skrih^{(t)}(f\1_{\{f>\eps\}})-\skrih(f\1_{\{f>
\eps\}})\bigr|-
(2R)^d\bigl|\widetilde H_t(\eps)-\widetilde H(\eps)\bigr|,
\end{multline}
where we invoked the explicit form of $f\mapsto\skrih(f)$. Since
both absolute values on the right-hand side tend to $0$ as
$t\to\infty$ uniformly in $f\le M$, the lower bound in
\eqref{ldpu} follows again by Varadhan's lemma and limit
$M\to\infty$. For the upper bound, the estimate and uniform limit
$\skrih^{(t)}(f)\le\skrih^{(t)}(f\1_{\{f>\eps\}})
\to\skrih(f\1_{\{f>\eps\}})$ give us a bound in terms of the
quantity $\chi_{3R}^\#(\eps)$ defined in \eqref{chicross}. By then
$M$ is irrelevant, so by invoking Proposition~\ref{chiprop}(3),
the claim is proved by taking $\eps\downarrow0$.

It remains to prove \eqref{ldplambda}. Recall the shorthand
$\lambda_k=\lambda^{{\rm d},k}_{R\alpha(t)}(\xi)$. By
\eqref{Fourierp}, \eqref{FKp} and analogously to \eqref{lowbound},
we have
\begin{equation}
\Bigl\langle \sum_{k} e^{t\lambda_k}\Bigr\rangle=
 \sum_{z\in Q_{R\alpha(t)}}\bigl\langle
p_{R\alpha(t)}(t,z,z)\bigr\rangle =
\Bigl\langle \sum_{z\in Q_{R\alpha(t)}}\E_z\Bigl[e^{(\xi,\ell_t)}
\1\{\tau_{R\alpha_t}>t\}\1\bigl\{X(t)=z\bigr\}\Bigr]\Bigr\rangle.
\end{equation}
Noting that $\1\{X(t)=z\}\leq 1$, we thus have $\langle \sum_{k}
e^{t\lambda_k}\rangle\leq \langle
(u_{R\alpha(t)}(t,\cdot),\1)\rangle$. With this in the hand,
\eqref{ldplambda} directly follows by the right inequality in
\eqref{ldpu}.
\end{proofsect}

\begin{proofsect}{Proof of Lemma~\ref{preduction}}
In the course of the proof, we use abbreviations $r=R\alpha(pt)$
and $\lambda_k=\lambda^{{\rm d},k} _{r}(\xi)$. Recall that $({\rm
e}_k)_{k}$ denotes an orthonormal basis in $\ell^2(Q_r)$ (with
inner product $(\cdot,\cdot)_r$) consisting of the
eigenfunctions of $\kappa\DeltaD+\xi$ with Dirichlet boundary condition.

We first turn to the case $p\geq 1$.
Use the Fourier expansion \eqref{Fourieru} and the inequality
\begin{equation}
\Bigl(\sum_{i=1}^n x_i\Bigr)^p\geq \sum_{i=1}^n x_i^p,\qquad
x_1,\ldots,x_n\geq 0,\,n\in\N,
\end{equation}
to obtain
\begin{equation}
\label{calc1}
\bigl\langle (u_r(t,\cdot),\1)^p\bigr\rangle =\Bigl\langle
\Bigl(\sum_{k} e^{t\lambda_k}
\,({\rm e}_k,\1)_r^2\Bigr)^p\Bigr\rangle\geq
\Bigl\langle \sum_{k} e^{pt\lambda_k}
\,({\rm e}_k,\1)_r^{2p}\Bigr\rangle.
\end{equation}
By Jensen's inequality for the probability measure
\begin{equation}
(l,d\xi)\mapsto\Bigl\langle\sum_{k}e^{pt\lambda_k}\Bigr\rangle^{-1}
e^{pt\lambda_l}\prob(d\xi),
\end{equation}
we have
\begin{equation}
\label{calc2}
\begin{aligned}
\quad\mbox{r.h.s.\ of (\ref{calc1}) } &\geq
\biggl(\frac{\langle\sum_{k} e^{pt\lambda_k} ({\rm
e}_k,\1)_r^2\rangle} {\langle\sum_{k}e^{pt\lambda_k}
\rangle}\biggr)^p \Bigl\langle\sum_{k} e^{pt\lambda_k}
\Bigr\rangle\\ &\geq e^{o(t\alpha_{pt}^{-2})} \Bigl\langle
\sum_{k}e^{pt\lambda_k}\,({\rm e}_k,\1)_r^2\Bigr\rangle =
e^{o(t\alpha_{pt}^{-2})}\bigl\langle
(u_r(pt,\cdot),\1)\bigr\rangle,
\end{aligned}
\end{equation}
where we recalled from the end of the proof of
Lemma~\ref{largedeviat} that
$\langle\sum_{k}e^{pt\lambda_k}\rangle\le\langle
(u_r(pt,\cdot),\1)\rangle =\langle\sum_{k} e^{pt\lambda_k} ({\rm
e}_k,\1)_r^2\rangle$, inserted $1\ge e^{o(t\alpha_{pt}^{-2})}({\rm
e}_k,\1)_r^2$, and applied \eqref{Fourieru}.

In the case $p\in(0,1)$, we apply Jensen's inequality as follows:
\begin{equation}
\bigl\langle (u_r(t,\cdot),\1)^p\bigr\rangle =
(\1,\1)_r^p\biggl\langle\Bigl(\sum_k e^{t\lambda_k}
\frac{({\rm e}_k,\1)_r^2}{(\1,\1)_r}\Bigr)^p\biggr\rangle
\geq (\1,\1)_r^p\biggl\langle\sum_k e^{pt\lambda_k}
\frac{({\rm e}_k,\1)_r^2}{(\1,\1)_r}\biggr\rangle.
\end{equation}
Invoking that $(\1,\1)_r=e^{o(t\alpha_{pt}^{-2})}$, the proof is finished
by recalling \eqref{Fourieru} once again.
\end{proofsect}

\subsection{The upper bound\label{momupper}}
Recall that $Q_R$ denotes the discrete box $[-R,R]^d\cap\Z^d$. We
abbreviate $r(t)=t\log t$ for $t>0$. For $z\in\Z^d$ and $R>0$, we
denote by $\lambdaD_{z;R}(V)$ the principal eigenvalue of the
operator $\kappa\DeltaD+V$ with Dirichlet boundary conditions in
the {\it shifted\/} box $z+Q_{R}$.
The main ingredient in the proof of the upper bound in
Theorem~\ref{momasy} is (the following) Proposition~\ref{compact},
which provides an estimate of $u(t,0)$ in terms of the maximal principal
eigenvalue of $\kappa\Delta^{\rm d}+V$ in small subboxes
(``microboxes'') of the ``macrobox'' $Q_{r(t)}$.

\begin{proposition}
\label{compact}
Let $B_R(t)=Q_{r(t)+2\lfloor R\rfloor}$. Then there
is a constant $C=C(d,\kappa)>0$ such that, for any $R, t>C$ and
any  potential $V\colon\Z^d\to [-\infty,0]$,
\begin{equation}\label{upperbound}
u^V(t,0)\leq e^{-t}+e^{Ct/R^2}\bigl(3r(t)\bigr)^d
\exp\left\{t\max_{z\in B_R(t)}\lambdaD_{z;2R}(V)\right\}.
\end{equation}
\end{proposition}

By Proposition~\ref{compact} and inequality \eqref{ldplambda}, the
upper bound in Theorem~\ref{momasy} is now easy:

\begin{proofsect}{Proof of Theorem~\ref{momasy}, upper bound}
Let $p\in (0,\infty)$.
First, notice that the second term in \eqref{upperbound}
can be estimated in terms of a sum:
\begin{equation}
\label{maxesti}
\exp\left\{t\max_{z\in B_R(t)} \lambdaD_{z;2R}(V)\right\}\leq
\sum_{z\in B_R(t)}e^{t\lambdaD_{z;2R}(V)}.
\end{equation}
Thus, applying \eqref{upperbound} to $u(t,0)$ (i.e., for $V=\xi$)
with $R$ replaced by $R\alpha(pt)$ for some fixed $R>0$,
raising both sides to the $p$-th power, and using \eqref{maxesti}
we get
\begin{equation}
u(t,0)^p\le 2^p\max\Bigl\{ e^{-pt},\,  e^{Cpt/(R^2\alpha(pt)^2)}
\bigl(3r(t)\bigr)^{pd}\!\!\!\sum_{z\in
B_{R\alpha(pt)}(t)}\!\!\!e^{pt\lambdaD_{z;2R\alpha(pt)}(\xi)}\Bigr\}.
\end{equation}
Next we take the expectation w.r.t.\ $\xi$ and note that, by
the shift-invariance of $\xi$,  the distribution of $\lambdaD_{z;
2R\alpha(pt)}(\xi)$ does not depend on $z\in\Z^d$. Take logarithm,
multiply by $\alpha_{pt}^2/(pt)$ and let $t\to\infty$. Then we
have that
\begin{equation}
\label{momldp}
\limsup_{t\to\infty}\frac {\alpha_{pt}^2}{pt}
\log\bigl\langle u(t,0)^p\bigr\rangle\leq
\frac C{R^2}+\limsup_{t\to\infty}\frac {\alpha_{pt}^2}{pt}\log\bigl\langle
\exp\{pt\lambdaD_{2R\alpha(pt)}(\xi)\}\bigr\rangle,
\end{equation}
where we also used that $e^{-pt}$, $r(t)^{pd}$, and
$\#B_{R\alpha(pt)}(t)$ are all $e^{o(t\alpha_{pt}^{-2})}$ as
$t\to\infty$. Since
\begin{equation}
\label{trivineq}
\exp\bigl\{pt \lambdaD_{R\alpha(pt)}(\xi)\bigr\}\leq \sum_{k} \exp\bigl\{pt
\lambda^{{\rm d},k}_{R\alpha(pt)}(\xi)\bigr\},
\end{equation}
\eqref{ldplambda} for $pt$ instead of $t$ implies that the second
term on the right-hand side of  \eqref{momldp} is bounded by
$-\chi_{6R}$. The upper bound in Theorem~\ref{momasy} then follows
by letting $R\to\infty$.
\end{proofsect}

Now we can turn to the proof of Proposition~\ref{compact}. We
begin by showing that $u^V(t,0)$ is very close to the solution
$u^V_{r(t)}(t,0)$ of the initial-boundary problem
\eqref{initialboundary}, whenever the size $r(t)=t\log t$ of the
``macrobox'' $Q_{r(t)}$ is large enough.

\begin{lemma}\label{box}
For sufficiently large $t>0$,
\begin{equation}\label{finitebox}
u^V(t,0)\leq e^{-t}+u^V_{r(t)}(t,0).
\end{equation}
\end{lemma}
\begin{proofsect}{Proof}
It is immediate from \eqref{FK} and \eqref{FKDir} with $r=r(t)$ that
\begin{equation}
\label{1.1}
u^V(t,0)-u^V_{r(t)}(t,0) =\E_0\left[
\exp\left\{\int_0^t V\bigl(X(s)\bigr)\,ds
\right\}\1\{\tau_{{r(t)}}\leq t\}\right].
\end{equation}
According to Lemma~2.5(a) in \cite{GM98}, we have, for every $r>0$,
\begin{equation}
\label{leaveabox}
\P_0(\tau_{r}\leq t)\leq 2^{d+1}\exp\left\{-r\left(\log\frac
r{d\kappa t}-1\right)\right\}.
\end{equation}
Using this for $r=r(t)=t\log t$ in \eqref{1.1}, we see that, for
sufficiently large $t$ (depending only on $d$ and $\kappa$),
the right-hand side of \eqref{1.1} is no more than $e^{-t}$.
\end{proofsect}
The crux of our proof of Proposition~\ref{compact} is that the
principal eigenvalue in a box $Q_r$ of size $r$ can be bounded by
the maximal principal eigenvalue in  ``microboxes'' $z+Q_R$
contained in $Q_r$, at the cost of changing the potential
slightly. This will later allow us to move the $t$-dependence of
the principal eigenvalue from the {\em size\/} of $Q_{r(t)}$ to
the {\em number\/} of ``microboxes.'' The following lemma is a
discrete version of Proposition~1 of \cite{GK98} and is based on
ideas from \cite{GM00}. However, for the sake of completeness, no
familiarity with \cite{GK98} is assumed.

\begin{lemma}
\label{eigencomp}
There is a number $C>0$ such that for every integer $R$, there
is a function $\Phi_R\colon\Z^d\to[0,\infty)$ with the
following properties:

(1) $\Phi_R$ is $2R$-periodic in every component.

(2) $\|\Phi_R\|_\infty\leq C/R^2$.

(3) For any potential $V\colon\Z^d\to[-\infty,0]$ and any $r>R$,
\begin{equation}\label{lambdaesti}
\lambdaD_{r}(V-\Phi_R)\leq\max_{z\in Q_{r+2R}}
\lambdaD_{z;{2R}}(V).
\end{equation}
\end{lemma}

\begin{proofsect}{Proof}
The idea is to construct a partition of unity
\begin{equation}
\label{partition}
\sum_{k\in\Z^d}\eta_k^2(z)=1,\qquad z\in\Z^d,
\end{equation}
where $\eta_k(z)=\eta(z-2Rk)$ with
\begin{equation}
\label{property}
\eta\colon\Z^d\to[0,1] \text{ such that }
\eta\equiv 1 \text{ on }Q_{R/2},\,\,
\supp(\eta)\subset Q_{3R/2}.
\end{equation}
Then we put
\begin{equation}\label{Phidef}
\Phi_R(z)=\kappa\sum_{k\in\Z^d}\bigl|\nablaD\eta_k(z)\bigr|^2,\qquad
z\in\Z^d,
\end{equation}
where $\nabla$ is the discrete gradient. Obviously, $\Phi_R$ is
$2R$-periodic in every component. The construction of $\eta$ such
that $\Phi_R$ satisfies (2) is given at the end of this proof.

Assuming the existence of the above partition of unity, we turn to
the proof of \eqref{lambdaesti}. Recall the Rayleigh-Ritz formula
\eqref{discreteeigenv}, which can be shortened as $\lambdaD_r(V)=\sup G^V(g)$,
where
\begin{equation}
G^V(g)=\sum_{z\in\Z^d} \bigl(-\kappa|\nablaD g(z)|^2+V(z)g^2(z)\bigr),
\end{equation}
and where the supremum is over normalized $g\in\ell^2(\Z^d)$ with
support in $Q_r$. Let $g$ be such a function, and define
$g_k(z)=g(z)\eta_k(z)$ for $k,z\in\Z^d$. Note that, according to
\eqref{partition} and \eqref{property}, we have
$\sum_k\|g_k\|_{2}^2=1$ and $\supp(g_k)\subset 2kR+Q_{3R/2}$. 

The pivotal point of the proof is the bound
\begin{equation}
\label{Gident} G^{V-\Phi_R}(g)\le\sum_{k\in\Z^d}\|g_k\|_{2}^2\,
G^V\Bigl(\frac{g_k} {\|g_k\|_{2}}\Bigr).
\end{equation}
In order to prove this inequality, we invoke the rewrite
\begin{equation}
g(y)\eta_k(y)-g(x)\eta_k(x)=g(x)\bigl(\eta_k(y)-\eta_k(x)\bigr)
+\eta_k(y)\bigl(g(y)-g(x)\bigr),
\end{equation}
recall \eqref{partition} and \eqref{Phidef}, and then perform a
couple of symmetrizations to derive
\begin{equation}
\label{helpeq} \kappa\sum_{k\in\Z^d}\sum_{x\in\Z^d}\bigl|\nabla
g_k(x)\bigr|^2= \sum_{x\in\Z^d}\Bigl[\kappa\bigl|\nabla
g(x)\bigr|^2+\Phi_R(x)g(x)^2\Bigr]+\kappa \Theta,
\end{equation}
where $\Theta$ is given by the formula
\begin{equation}
\Theta=-\frac12\sum_{k\in\Z^d}\sum_{x\in\Z^d} \sum_{y\colon\!
y\sim x}\bigl[g(y)-g(x)\bigr]^2
\bigl[\eta_k(y)-\eta_k(x)\bigr]^2\le0.
\end{equation}
Using this bound on the right-hand side of \eqref{helpeq}, we have
\begin{equation}
\begin{aligned}
\sum_{k\in\Z^d}\|g_k\|_{2}^2\, G^V\Bigl(\frac{g_k}
{\|g_k\|_{2}}\Bigr) &=\sum_{k\in\Z^d}G^V(g_k)
=\sum_{z\in\Z^d}\sum_{k\in\Z^d}\Bigl[-\kappa\bigl|\nablaD
g_k(z)\bigr|^2+V(z)g_k^2(z)\Bigr]\\
&\ge\sum_{z\in\Z^d}\Bigl[-\kappa \bigl|\nablaD g(z)\bigr|^2+
\bigl(V(z)-\Phi_R(z)\bigr)g^2(z)\Bigr] =G^{V-\Phi_R}(g),
\end{aligned}
\end{equation}
which is exactly the inequality \eqref{Gident}.

Since the support of $g_k$ is contained in $2kR+Q_{3R/2}$, the
Rayleigh-Ritz formula yields that 
\begin{equation}
G^V\Bigl(\frac{g_k}
{\|g_k\|_{2}}\Bigr)\leq
\lambdaD_{2kR;{3R/2}}(V)\le \lambdaD_{2kR;{2R}}(V)
\end{equation} 
whenever $\|g_k\|_{2}\not=0$ (which requires, in particular, that 
$2R|k|-3R/2\leq r$). Estimating these eigenvalues by their 
maximum and taking into account that $\sum_{k\in\Z^d}\|g_k\|_{2}^ 
2=\|g\|_{2}^2=1$, we find that the right-hand side of 
\eqref{Gident} does not exceed the right-hand side of 
(\ref{lambdaesti}). The claim~\eqref{lambdaesti} is finished by 
passing to the supremum over $g$ on the left-hand side of 
\eqref{Gident}.

For the proof to be complete, it remains to construct the
functions $\eta$ and $\Phi_R$ with the properties
\eqref{partition} and \eqref{property} and such that
$\|\Phi_R\|_\infty\leq C/R^2$ for some $C>0$. First, the ansatz
\begin{equation}
\eta(z)=\prod_{i=1}^d \zeta(z_i),\qquad z=(z_1,\ldots,z_d)\in\Z^d,
\end{equation}
reduces the construction of $\eta$ to the case $d=1$ (with $\eta$
replaced by $\zeta$). In order to define $z\mapsto\zeta(z)$, let
$\varphi\colon\R\to[0,1]$ be such that both $\sqrt\varphi$ and
$\sqrt{1-\varphi}$ are smooth, $\varphi\equiv 0$ on $(-\infty,-1]$
and $\varphi\equiv 1$ on $[0,\infty)$ and
$\varphi(-x)=1-\varphi(x)$ for all $x\in\R$. Then we put
\begin{equation}
\textstyle \zeta(z)=\sqrt{\varphi\bigl(\frac12+\frac
zR\bigr)\bigl[ 1-\varphi\bigl(-\frac32+\frac
zR\bigr)\bigr]},\qquad z\in\Z.
\end{equation}
In order to verify that the functions
$\zeta_k^2(z)=\zeta^2(z+2Rk)$ with $k\in\Z$ form a partition of
unity on $\R$, we first note that $\zeta(z)\equiv 1$ on
$[-R/2,R/2]$ while
$\zeta(z)+\zeta(z-2R)=1-\varphi(-3/2+z/R)+\varphi(-3/2+z/R)=1$ for
$z\in[R/2,3R/2]$. Moreover, as follows by a direct computation,
$\sup_{z\in\Z}\sum_k|\nabla\zeta_k(z)|^2\le
4\Vert(\sqrt\varphi)^\prime\Vert_\infty^2 R^{-2}$, which means
that (2) is satisfied with
$C=4d\Vert(\sqrt\varphi)^\prime\Vert_\infty^2$. This finishes the
construction and also the proof. 
\end{proofsect}

\begin{proofsect}{Proof of Proposition~\ref{compact}}
Having all the prerequisites, the proof is easily completed.
First,
\begin{equation}
\int_0^t V\bigl(X(s)\bigr)\,ds\leq t\frac
C{R^2}+\int_0^t(V-\Phi_R)\bigl(X(s)\bigr)\,ds,\qquad t>0.
\end{equation}
by Lemma~\ref{eigencomp}(2). Therefore, combining \eqref{FK} with
Lemma~\ref{box}, we have that
\begin{equation}
u^V(t,0)\leq e^{-t}+e^{tC/R^2}u^{V-\Phi_R}_{{r(t)}}(t,0)
\end{equation}
whenever $t$ is large enough.
Invoking also the Fourier expansion \eqref{Fourieru} w.r.t.\ the
eigenfunctions of $\kappa\DeltaD+V-\Phi_R$ in $\ell^2(Q_{r(t)})$
and the fact that $(\1,\1)_{r(t)}=\# Q_{r(t)}$,
we find that
\begin{equation}
u^{V-\Phi_R}_{{r(t)}}(t,0)\leq \sum_{z\in Q_{r(t)}}u^{V-\Phi_R}_{{r(t)}}(t,z)
\leq \# Q_{r(t)}\,\exp\bigl\{t\lambdaD_{{r(t)}}(V-\Phi_R)\bigr\}.
\end{equation}
Now apply Lemma~\ref{eigencomp} for $r=r(t)=t\log t$ to finish the
proof.
\end{proofsect}
\subsection{Proof of Lifshitz tails\label{pfofLifshitz}}
Let $\nu_R$ denote the empirical measure on the spectrum of
${\mathfrak H}_R$, i.e.,
\begin{equation}
\nu_R=\frac1{\# Q_R}\sum_k\delta_{\{-\lambda_k\}},
\end{equation}
where $\lambda_k=\lambda_{R}^{{\rm d},k}(\xi)=-E_k$ denotes the
eigenvalues of $-{\mathfrak H}_R$. Note that $\nu_R$ has total
mass at most $1$, because the dimension of the underlying Hilbert
space is bounded by $\#Q_R$. Due to \eqref{mainass}, $\nu_R$ is
supported on $[0,\infty)$. Moreover, $N_R(E)$ in
\eqref{numberofstates} is precisely $\#Q_R\,\nu_R([0,E])$, for any
$E\in[0,\infty)$. Let ${\mathcal L}(\nu_R,t)$ be the Laplace
transform of $\nu_R$ evaluated at $t\ge 0$,
\begin{equation}
\label{Lapltr}
{\mathcal L}(\nu_R,t)=\int \nu_R(d\lambda)\, e^{-\lambda
t}=\frac1{\# Q_R}\sum_k e^{t\lambda_k}.
\end{equation}
Adapting Theorem~VI.1.1.\ in \cite{CL90} to our discrete setting,
the existence of the limit \eqref{IDSlimit} is proved by
establishing the a.s.\ convergence of $\nu_R$ to some non-random
$\nu$, which in turn is done by proving that ${\mathcal
L}(\nu_R,\cdot)$ has a.s.\ a non-random limit. In our case, the
argument is so short that we find it convenient to reproduce it
here.

Invoking \eqref{Fourierp} and \eqref{FKp} for $V=\xi$, we
have from \eqref{Lapltr} that
\begin{equation}
\label{calLnuR}
{\mathcal L}(\nu_R,t)=\frac1{\# Q_R}\sum_{z\in Q_R}{\mathbb
E}_z\Bigl\{\exp\Bigl[ \int_0^t\xi\bigl(X(s)\bigr)ds\Bigr]
\1\{\tau_R>t\}\1\bigl\{X(t)=z\bigr\}\Bigr\}.
\end{equation}
Next, writing $\1\{\tau_R>t\}=1-\1\{\tau_R\le t\}$ we arrive at
two terms, the second of which tends to zero as $R\to\infty$ for
any fixed $t$ by the estimate
\begin{equation}
0\le\frac1{\# Q_R}\sum_{z\in Q_R}{\mathbb E}_z\Bigl\{e^{
\int_0^t\xi(X(s))ds}\1\{\tau_R\le t\}\1\{X(t)=z\}\Bigr\}\le
\frac1{\# Q_R}\sum_{z\in Q_R}{\mathbb P}_z(\tau_R\le t),
\end{equation}
where we used that $\xi\le 0$. Indeed, ${\mathbb P}_z(\tau_R\le
t)\le {\mathbb P}_0(\tau_{R(z)}\le t)$ with
$R(z)=\operatorname{dist}(z,Q_R^{\rm c})$, which by
\eqref{leaveabox} means that ${\mathbb P}_z(\tau_R\le t)$ decays
exponentially with $\operatorname{dist}(z,Q_R^{\rm c})$. Thus,
${\mathcal L}(\nu_R,t)$ is asymptotically given by the right-hand
side of \eqref{calLnuR} with $\1\{\tau_R>t\}$ omitted. But then
the right-hand side is the average of an $L^1$ function over the
translates in the box $Q_R$, so by the Ergodic Theorem,
\begin{equation}
\label{pfoflimit}
\lim_{R\to\infty}{\mathcal L}(\nu_R,t)=\biggl\langle{\mathbb
E}_0\Bigl\{\exp\Bigl[
\int_0^t\xi\bigl(X(s)\bigr)ds\Bigr]\1\bigl\{X(t)=0\bigr\}
\Bigr\}\biggr\rangle
\end{equation}
$\xi$-almost surely for every fixed $t\ge 0$ (the exceptional null
set is {\it a priori\/} $t$-dependent). Both the right-hand
side of \eqref{pfoflimit} and ${\mathcal L}(\nu_R,t)$ for every
$R$ are continuous and decreasing in $t$. Consequently, with
probability one \eqref{pfoflimit} holds for all $t\ge0$.

The right-hand side of \eqref{pfoflimit} inherits the complete
monotonicity property from ${\mathcal L}(\nu_R,t)$; it thus equals
${\mathcal L}(\nu,t)$ where $\nu$ is some measure supported in
$[0,\infty)$. Moreover, this also implies that $\nu_R\to\nu$
weakly as $R\to\infty$. In particular, we have $n(E)=\nu([0,E])$
for any $E\geq 0$.

\begin{proofsect}{Proof of Theorem~\ref{Lifshitz}}
From \eqref{pfoflimit} we immediately have
\begin{equation}
\label{sandwich}
e^{o(t/\alpha_t^2)}\bigl\langle e^{t\lambdaD_{R\alpha(t)}}\bigr\rangle
\le {\mathcal L}(\nu,t)\le\langle u(t,0)\rangle,\qquad R\ge 0,
\end{equation}
where $\lambdaD_{R\alpha(t)}$ is as in \eqref{discreteeigenv}.
Here, for the upper bound we simply neglected $\1\{X(t)=0\}$ in
\eqref{pfoflimit}, whereas for the lower bound we first wrote
\eqref{pfoflimit} as a normalized sum of the right-hand side of
\eqref{pfoflimit} with the walk starting and ending at all
possible $z\in Q_{R\alpha_t}$, and then inserted $\1\{\supp
(\ell_t)\subset Q_{R\alpha(t)}\}$, applied \eqref{FKp} and
\eqref{Fourierp}, and then recalled \eqref{trivineq}. The factor
$e^{o(t/\alpha_t^2)}$ comes from the normalization by $\#
Q_{R\alpha(t)}$ in the first step. Using subsequently
\eqref{momldp} for $p=1$, the left-hand side of \eqref{sandwich}
is further bounded from below by
$e^{(t/\alpha_t^2)(-4C/R^2+o(1))}\langle u(t,0)\rangle$. Then
Theorem~\ref{momasy} and the limit $R\to\infty$ enable us to
conclude that
\begin{equation}
\label{limlogL}
\lim_{t\to\infty} \frac{\alpha_t^2}t \log {\mathcal
L}(\nu,t)=-\chi.
\end{equation}

In the remainder of the proof, we have to convert this statement
into the appropriate limit for the IDS. This is a standard problem
in the theory of Laplace transforms and, indeed, there are
theorems that can after some work be applied (e.g., de Bruijn's
Tauberian Theorem, see Bingham, Goldie and Teugels \cite{BGT87}).
However, for the sake of both completeness and convenience we
provide an independent proof below.

Suppose that $H$ is the $\gamma$-class. We begin with an upper
bound. Clearly,
\begin{equation}
{\mathcal L}(\nu,t)\ge e^{-t E} n(E) \quad\text{for any }t, E\ge0.
\end{equation}
Let $t_E=\alpha^{-1}(\sqrt{(1-2\nu)\chi \,E^{-1}})$ and insert
this for $t$ in the previous expression. The result is
\begin{equation}
\label{upper}
\log n(E)\le t_E E+ \log{\mathcal L}(\nu,t_E)= -t_E
E\textstyle\frac{2\nu}{1-2\nu}\bigl(1+o(1)\bigr),\qquad
E\downarrow 0,
\end{equation}
where we applied \eqref{limlogL} and the definition of $t_E$. In
order to finish the upper bound, we first remark that from the
first assertion in \eqref{scalingexp1} it can be deduced that
\begin{equation}
\label{ttoE}
\lim_{E\downarrow0}\frac{t_E}{\alpha^{-1}(E^{-\frac12})}
=\bigl[(1-2\nu)\chi]^{-\frac1{2\nu}}.
\end{equation}
Indeed, define $t_E^\prime=\alpha^{-1}(E^{-1/2})$ and consider the
quantity $p_E=t_E/t_E^\prime$. Clearly,
\begin{equation}
\label{p_E}
\alpha(p_E t_E^\prime)=\alpha(t_E^\prime)\sqrt{(1-2\nu)\chi}.
\end{equation}
Let $\widetilde p=[(1-2\nu)\chi]^{-1/(2\nu)}$. Since
$t_E^\prime\to\infty$ as $E\downarrow0$, there is no $\eps>0$ such
that $p_E\ge\widetilde p+\eps$ for infinitely many $E$ with an
accumulation point at zero, because otherwise the left-hand side
\eqref{p_E} would, by \eqref{scalingexp1}, eventually exceed the
right-hand side. Similarly we prove that
$\liminf_{E\downarrow0}p_E$ cannot be smaller than $\widetilde
p-\eps$. Therefore, $p_E\to\widetilde p$ as $E\downarrow0$, which
is \eqref{ttoE}.

Using \eqref{ttoE}, we have from \eqref{upper} that
\begin{equation}
\limsup_{E\downarrow0}\frac{\log n(E)}{E\alpha^{-1}(E^{-\frac12})}\le
-\frac{2\nu}{1-2\nu}\bigl[(1-2\nu)\chi]^{-\frac1{2\nu}}.
\end{equation}

The lower bound is slightly harder, but quite standard. First,
introduce the probability measure on $[0,\infty)$ defined by
\begin{equation}
\mu_E(d\lambda)=\frac{e^{-t_E\lambda}}{{\mathcal
L}(\nu,t_E)}\nu(d\lambda),\quad E\ge0.
\end{equation}
We claim that, for any $\eps>0$, all mass of $\mu_E$ gets
eventually concentrated inside the interval $[E-\eps E,E+\eps E]$
as $E\downarrow 0$. Indeed, for any $0\le t<t_E$ we have
\begin{equation}
\mu_E\bigl((E+\eps E,\infty)\bigr)\le {\mathcal
L}(\nu,t_E)^{-1}\int_{E+\eps E}^\infty \nu(d\lambda)\,
e^{-t_E\lambda+t(\lambda-E-\eps E)}\le
e^{-t\eps E}\frac{{\mathcal L}(\nu,t_E-t)}{{\mathcal
L}(\nu,t_E)}e^{-tE}.
\end{equation}
Pick $0<\delta<1$ and set $t=\delta t_E$. Then we have
\begin{equation}
\mu_E\bigl((E+\eps E,\infty)\bigr)\le\exp\Bigl\{-\delta\eps
t_EE-\delta t_EE-\chi \textstyle \frac{t_E}{\alpha(t_E)^2}\bigl[
(1-\delta)^{1-2\nu}-1+o(1)\bigr]\Bigr\},
\end{equation}
where we again used \eqref{limlogL} and \eqref{scalingexp1}.
Applying that $(1-\delta)^{1-2\nu}-1=-\delta(1-2\nu)+o(\delta)$,
using
\begin{equation}
t_E E-\chi (1-2\nu)\textstyle \frac{t_E}{\alpha(t_E)^2}=0,
\end{equation}
and noting that $\alpha(t_E)^{-2}=O(E)$, we have
\begin{equation}
\label{finupper}
\mu_E\bigl((E+\eps E,\infty)\bigr)\le \exp \bigl[-t_E
E\bigl(\delta\eps+o(\delta)\bigr)\bigr].
\end{equation}
Choosing $\delta$ small enough, the right-hand side vanishes as
$E\downarrow0$. Similarly we proceed in the case $[0,E-\eps E)$.

Now we can finish the lower bound on Lifshitz tails. Indeed, using
Jensen's inequality
\begin{equation}
\label{almostfin}
\begin{aligned}
\nu\bigl([0,E+\eps E]\bigr)&={\mathcal L}(\nu,t_E) \int_0^{E+\eps
E} \!\!\!\mu_E(d\lambda) \,e^{t_E\lambda}\\ &\ge {\mathcal
L}(\nu,t_E)\mu_E\bigl([0,E+\eps E]\bigr) \exp\Bigl\{
{\textstyle\frac{t_E}{\mu_E([0,E+\eps E])}}\int_0^{E+\eps E}
\!\!\!\!\!\mu_E(d\lambda)\,\lambda\Bigr\}.
\end{aligned}
\end{equation}
But $\int_0^\infty \mu_E(d\lambda)\lambda$ tends to $E$, by what
we have proved about the concentration of the mass of $\mu_E$
(note that \eqref{finupper} and the similar bound for $[0,E-\eps
E)$ are both exponential in $\eps$) and, by the same token, so
does $\int_0^{E+\eps E} \mu_E(d\lambda)\lambda$. By putting all
this together, dividing both sides of \eqref{almostfin} by
$E^\prime\alpha^{-1}((E^\prime)^{-1/2})$ with $E^\prime=E+\eps E$,
interpreting $E^\prime$ as a new variable tending to $0$ as
$E\downarrow0$, and invoking \eqref{upper} and the subsequent
computation, we get
\begin{equation}
\liminf_{E\downarrow0}\frac{\log
n(E)}{E\alpha^{-1}(E^{-\frac12})}\ge
-(1+\eps)^{\frac{1-2\nu}{2\nu}}
\frac{2\nu}{1-2\nu}\bigl[(1-2\nu)\chi\bigr]^{-\frac1{2\nu}},
\end{equation}
where we also used that $t_E/t_{E+\eps E}\to
(1+\eps)^{1/(2\nu)}$. Since
$\eps$ was arbitrary, the claim is finished by taking
$\eps\to0$. \end{proofsect}

\section{Proof of Theorem~\ref{asasy}}\label{asproof}\noindent
Again, we divide the proof in two parts: the upper bound and the 
lower bound. While the former is a simple application of our 
results on the moment asymptotics (and the exponential Chebyshev
inequality), the latter requires two ingredients: a 
Borel-Cantelli argument for size of the field and a rather 
tedious percolation argument. These combine in 
Proposition~\ref{ximicrobox}, whose proof is deferred to 
Subsection~\ref{tech-claims}.

\subsection{The upper bound}
\vspace{-6mm}
\begin{proofsect}{Proof of Theorem~\ref{asasy}, upper bound}
Let $r(t)=t\log t$ and let $L\in(0,\infty)$. We want to apply
Proposition~\ref{compact} with the random potential $V=\xi$ and
with $R$ replaced by $R\alpha(Lb_t)$ for some fixed $R,L>0$.
(Later we shall let $R\to\infty$ and pick $L$ appropriately.)

Recall the definition of $B_R(t)$ in Proposition~\ref{compact} and
abbreviate $B(t)=B_{R\alpha(Lb_t)}(t)$. Take logarithms in
\eqref{upperbound}, multiply by $\alpha_{b_t}^2/t$ and use
\eqref{scalingexp1} to obtain
\begin{equation}\label{upperas}
\limsup_{t\to\infty}\frac {\alpha_{b_t}^2}t\log u(t,0)\leq \frac
C{L^{2\nu} R^2}+\limsup_{t\to\infty}\Bigl[\alpha_{b_t}^2
\max_{z\in B(t)}\lambdaD_{z;{2R\alpha(Lb_t)}}(\xi)\Bigr],
\end{equation}
almost surely w.r.t.\ the field $\xi$. Thus, we just need to
evaluate the almost sure behavior of the maximum of the random
variables on the right-hand side. This will be done by showing
that
\begin{equation}\label{maxbound}
\limsup_{R\to\infty}\limsup_{t\to\infty}\Bigl[\alpha_{b_t}^2
\max_{z\in B(t)}\lambdaD_{z;{2R\alpha(Lb_t)}}(\xi)\Bigr]\leq
-\widetilde\chi
\end{equation}
almost surely w.r.t.\ the field $\xi$, provided $L>0$ is chosen
appropriately.

For any $t>0$, let $(\lambda_i(t))_{i=1,\dots,N(t)}$ be an
enumeration of the random variables
$\lambdaD_{z;{2R\alpha(Lb_t)}}(\xi)$ with $z\in B(t)$. Note
that  $N(t)\le3^d t^d(\log t)^d$ for $t$ large. Clearly,
$(\lambda_i(t))$ are identically distributed but not independent.
By \eqref{ldplambda}, the tail of their distribution is bounded by
\begin{equation}
\label{outline3}
\limsup_{t\to\infty}\frac{\alpha_{b_t}^2}{b_t} \log\bigl\langle
\exp\{Lb_t \lambdaD_{{2R\alpha(Lb_t)}} (\xi)\}\bigr\rangle\leq-
L^{1-2\nu}\chi_{6R},\qquad L,R>0,
\end{equation}
where $\chi_R$ is defined in \eqref{chiRdef}.

The assertion \eqref{maxbound} will be proved if we can verify
that, with probability~one,
\begin{equation}\label{maximum}
\max_{i=1,\dots,N(t)}\lambda_i(t)\leq  -\frac{\widetilde
\chi-\eps}{\alpha^2(b_t)} \bigl(1+o(1)\bigr),\quad t\to\infty,
\end{equation}
for any $\eps>0$ and sufficiently large $R>0$, as $t\to\infty$. To
that end, note first that the left-hand side of \eqref{maximum} is
increasing in $t$ since the maps $t\mapsto \alpha(Lb_t)$,
$R\mapsto \lambda_R^{\rm d}(\xi)$ and $t\mapsto r(t)$ are all
increasing. As a consequence, it suffices to prove the assertion
\eqref{maximum} only for $t\in\{e^n\colon n\in\N\}$, because also
$\alpha(b_s)^{-2}- \alpha(b_{e^{n}})^{-2}=o(\alpha(b_{e^n})^{-2})$
as $n\to\infty$ for any $e^{n-1}\le s<e^n$. Let
\begin{equation}
p_n =\prob\left(\max_{i=1,\dots,N(e^n)} \lambda_i(e^n) \geq
-\frac{\widetilde\chi-\eps}{\alpha^2(b_{e^n})}\right).
\end{equation}
Abbreviating $t=e^n$ and recalling $b_t\alpha_{b_t}^{-2}=\log
t=n$, the exponential Chebyshev inequality and \eqref{outline3}
allow us to write for any $L>0$ and $n$ large that
\begin{equation}
\label{outline5}
\begin{aligned}
p_n\le N(e^n)\,\prob\Bigl(&e^{L b_t\lambda_1(e^n)}\geq e^{-L b_t
\alpha^{-2}(b_t)(\widetilde\chi-\eps)}\Bigr)\\ \leq 3^d &n^d
e^{nd}\exp\bigl\{L b_t \alpha^{-2}(b_t)(\widetilde
\chi-\eps)\bigr\}\Bigl\langle e^{L
b_t\lambdaD_{2R\alpha(Lb_t)}(\xi)}\Bigr\rangle\\ &\qquad\qquad=
\exp\Bigl\{n\bigl[-\eps L+d+L\widetilde\chi-L^{1-2\nu}
\chi_{6R}+o(1)\bigl]\Bigr\}.
\end{aligned}
\end{equation}
Now let $L$ to minimize the function $L\mapsto 
d+L\widetilde\chi-L^{1-2\nu} \chi$ on $[0,\infty]$. An easy 
calculation reveals that $L=[(1-2\nu)\chi/ 
\widetilde\chi\,]^{1/(2\nu)}$. By invoking 
Proposition~\ref{tildechiident}, we also find that 
$d+L\widetilde\chi-L^{1-2\nu} \chi=0$ for this value of $L$, and, 
substituting this into \eqref{outline5}, we obtain
\begin{equation}
p_n\leq \exp\bigl\{-n\bigl[\eps
L-L^{1-2\nu}(\chi-\chi_{6R})+o(1)\bigr]\bigr\},
\end{equation}
which is clearly summable on $n$ provided $R$ is sufficiently
large. The Borel-Cantelli lemma then guarantees the validity of
\eqref{maximum}, which in turn proves \eqref{maxbound}. The limit
$R\to\infty$ then yields the upper bound in Theorem~\ref{asasy}.
\end{proofsect}

\subsection{The lower bound}
Recall the notation of Subsection~\ref{FKform}. Let
$Q_{\gamma_t}=[-\gamma_t,\gamma_t]^d\cap\Z^d$ denote the
``macrobox,'' where $\gamma_t$ is the time scale defined by
\begin{equation}\label{gammat}
\gamma_t=\frac t{\alpha_{b_t}^{3}},\qquad t>0.
\end{equation}
We assume without loss of generality that $t\mapsto\gamma_t$ is
strictly increasing. Since we assumed
$\prob(\xi(0)>-\infty)>p_{\rm c}(d)$ for $d\geq2$, there is a
$K\in(0,\infty)$ such that $\prob(\xi(0)\ge-K)>p_{\rm c}(d)$.
Consequently, $\{z\in\Z^d\colon \xi(z)\ge-K\}$ contains
almost-surely a unique infinite cluster ${\mathcal C}^*_\infty$.

Given a $\psi\in C^-([-R,R]^d)$, let
$\psi_t\colon\Z^d\to(-\infty,0]$ be the function
$\psi_t(\cdot)=\psi(\cdot/\alpha(b_t))/\alpha(b_t)^2$. Suppose $H$
is in the $\gamma$-class. Abbreviate
\begin{equation}
\label{Qdef}
Q^{(t)}=\begin{cases}Q_{R\alpha(b_t)}\quad&\text{if
}\gamma\not=0,\\
Q_{R\alpha(b_t)}\cap\supp\psi_t\quad&\text{if }\gamma=0.\\
\end{cases}
\end{equation}
The main point of the proof of the lower bound in
Theorem~\ref{asasy} is the existence of a microbox of diameter of
order $\alpha_{b_t}$ in $Q_{\gamma_t}$ (which is contained in
${\mathcal C}^*_\infty$ for $d\geq 2$) where the field is bounded
from below by $\psi_t$:

\begin{proposition}\label{ximicrobox}
Let $R>0$ and fix a function $\psi\in C^-(R)$ satisfying
$\L_R(\psi)< d$. Let $\eps>0$ and let $H$ be in the $\gamma$-class
with $\gamma\in[0,1)$. Then the following holds almost
surely: There is a\/ $t_0=t_0(\xi,\psi,\eps,R)<\infty$ such that
for each $t\ge t_0$, there exists a\/ $y_t\in Q_{\gamma_t}$ such
that
\begin{equation}\label{xilowerbound}
\xi(z+y_t)\geq\frac 1{\alpha_{b_t}^2}\psi\left(\frac
{z}{\alpha_{b_t}}\right) -\frac \eps{\alpha_{b_t}^2} \qquad
\forall z\in Q^{(t)}.
\end{equation}
In addition, whenever $d\ge 2$, $y_t$ can be chosen such that
$y_t\in{\mathcal C}^*_\infty$.
\end{proposition}

The proof of Proposition~\ref{ximicrobox} is deferred to
Subsection~\ref{tech-claims}. In order to make use of it, we need
that the walk can get to $y_t+Q^{(t)}$ in a reasonable time. In
$d\ge 2$, this will be possible whenever the above microbox can be
reached from any point in ${\mathcal C}_\infty^*\cap Q_{\gamma_t}$
by a path in ${\mathcal C}_\infty^*$ whose length is comparable to
the lattice distance between the path's end-points. Given
$x,z\in{\mathcal C}_\infty^*$, let $\operatorname{d}_*(x,z)$
denote the length of the shortest path in ${\mathcal C}_\infty^*$
connecting $x$ and $z$. Let $|x-z|_1$ be the lattice distance of
$x$ and $z$. The following lemma is the site-percolation version
of Lemma~2.4 in Antal's thesis \cite{A94}, page~72. While the
proof is given there in the bond-percolation setting, its
inspection shows that it carries over to our case. Therefore, we
omit it.

\begin{lemma}
\label{d^*/d} Suppose $d\ge 2$. Then, with probability one,
\begin{equation}\label{cluster}
\varrho(x):=\sup_{z\in{\mathcal C}_\infty^*\setminus\{x\}}
\frac{\operatorname{d}_*(x,z)}{|x-z|_1}<\infty\quad\text{ for all
}x\in {\mathcal C}_\infty^*.
\end{equation}
\end{lemma}

We proceed with the proof of Theorem~\ref{asasy} in the case $d\ge
2$. In $d=1$, Lemma~\ref{d^*/d} will be substituted by a different
argument.

\begin{proofsect}{Proof of Theorem~\ref{asasy} ($d\ge2$), lower bound}
Let $R,\eps>0$ and let $\psi\in C^-(R)$ be twice continuously
differentiable with $\L_R(\psi)< d$. If $\gamma=0$, let
$\supp\psi$ be a non-degenerate ball in $Q_R$ centered at $0$.
Suppose that $\xi=(\xi(z))_{z\in\Z^d}$ does not belong to the
exceptional null sets of the preceding assertions. In particular,
there are unique infinite clusters ${\mathcal C}_\infty$ in
$\{z\in\Z^d\colon\xi(z)>-\infty\}$ and ${\mathcal C}_\infty^*$ in
$\{z\in\Z^d\colon \xi(z)\ge -K\}$, and $\xi$ satisfies the claims
in Proposition~\ref{ximicrobox} and Lemma~\ref{d^*/d}. Clearly,
${\mathcal C}_\infty^*\subset{\mathcal C}_\infty$. Assume
$0\in{\mathcal C}_\infty$ and pick a $z^*\in{\mathcal
C}_\infty^*$. For each $t\ge t_0$ choose a $y_t\in
Q_{\gamma_t}\cap{\mathcal C}_\infty^*$ such that
\eqref{xilowerbound} holds. We assume that $t$ is so large that
$z^*\in Q_{\gamma_t}$.

The lower bound on $u(t,0)$ will be obtained by restricting the
random walk $(X(s))_{s\geq 0}$ (which starts at 0) to be at $z^*$ at
time $1$, at $y_t$ at time $\gamma_t$ (staying within ${\mathcal
C}_\infty^*$ in the meantime) and to remain in
$y_t+Q^{(t)}$ until time $t$. Introduce the exit times
from ${\mathcal C}_\infty^*$ and $y_t+Q^{(t)}$,
respectively,
\begin{equation}
\tau_\infty^*=\inf\bigl\{s>0\colon X(s)\notin{\mathcal C}_\infty^*\bigr\}
\qquad\mbox{and}\qquad
\tau_{y_t,t}=\inf\bigl\{s>0\colon X(s)\notin {y_t}+Q^{(t)}\bigr\}.
\end{equation}
Let $t\ge t_0(\xi)$. Inserting the indicator on the event described
above and using the Markov property twice at times $1$ and
$\gamma_t$, we get
\begin{equation}
\label{123}
u(t,0)\geq \text{I}\times \text{II}\times \text{III},
\end{equation}
where the three factors are given by
\begin{equation}\label{IIIIII}
\begin{aligned}
\text{I}&=\E_0\Bigl[\exp\Bigl\{\int_0^1
\xi\bigl(X(s)\bigr)\,ds\Bigr\}\1\bigl\{X(1)=z^*\bigr\}\Bigr],\\
\text{II}&=\E_{z^*}\Bigl[\exp\Bigl\{\int_0^{\gamma_t-1}\xi
\bigl(X(s)\bigr)\,ds\Bigr\}
\1\bigl\{\tau_\infty^*>\gamma_t-1,X(\gamma_t-1)=y_t\bigr\}\Bigr],\\
\text{III}&=\E_{y_t}\Bigl[\exp\Bigl\{\int_0^{t-\gamma_t}
\xi\bigl(X(s)\bigr)\,ds\Bigr\}\1\bigl\{\tau_{{y_t},t}>t-\gamma_t
\bigr\}\Bigr].
\end{aligned}
\end{equation}

Clearly, the quantity $\text{I}$ is independent of $t$ and is
non-vanishing because $0,z^*\in{\mathcal C}_\infty$. Our next
claim is that $\text{II}\geq e^{o(t\alpha_{b_t}^{-2})}$ as
$t\to\infty$. Indeed,
\begin{equation}
\text{II}\geq e^{-K
\gamma_t}\P_{z^*}\bigl(\tau_\infty^*>\gamma_t-1,X(\gamma_t-1)={y_t}\bigr),
\end{equation}
since there is at least one path connecting $z^*$ to $y_t$ within
${\mathcal C}_\infty^*$ (recall that the field $\xi$ is bounded
from below by $-K$ on ${\mathcal C}_\infty^*$). Denote by
$\operatorname{d}_t=\operatorname{d}_*(z^*,y_t)$ the minimal
length of such a path and abbreviate $\varrho(z^*)=\varrho$, where
$\varrho(z^*)$ is as in \eqref{cluster}. Then, for $t\ge t_0$,
\begin{equation}\label{dnesti}
\operatorname{d}_t\leq \varrho |z^*-y_t|_1\leq 2d\varrho
\gamma_t\leq 3d\varrho (\gamma_t-1),
\end{equation}
by Lemma~\ref{d^*/d} and the fact that the both $z^*,y_t\in
Q_{\gamma_t}$. Hence, using also that $\operatorname{d}_t!\leq
\operatorname{d}_t^{\operatorname{d}_t}$,
\begin{multline}
\label{prob1}
\P_{z^*}\bigl(\tau_\infty^*>\gamma_t-1,X(\gamma_t-1)=y_t\bigr)
\geq e^{-(\gamma_t-1)}\frac{(\gamma_t-1)^{\operatorname{d}_t}}
{\operatorname{d}_t!} (2d)^{-\operatorname{d}_t}\\ \geq
e^{-\gamma_t}\exp\bigl[-\operatorname{d}_t\log(2d
\operatorname{d}_t/(\gamma_t-1))\bigr] \geq
\exp\bigl[-\gamma_t\bigl(1+3d\varrho\log(6d^2\varrho)\bigr)\bigr].
\end{multline}
In order to see that $\text{II}\geq e^{o(t\alpha_{b_t}^{-2})}$,
recall that  $\gamma_t=o(t\alpha_{b_t}^{-2})$ as $t\to\infty$ by
\eqref{gammat} and that $z^*$ does not depend on $t$.

We turn to the estimate of $\text{III}$. By spatial homogeniety of
the random walk, we have
\begin{equation}
\text{III}=\E_0\Bigl[\exp\Bigl\{\int_0^{t-\gamma_t}\xi\bigl(y_t+X(s)\bigr)
\,ds\Bigr\}\1\{\tau_{0,t}>t-\gamma_t\}\Bigr],
\end{equation}
where $\tau_{0,t}$ is the first exit time from $Q^{(t)}$. Using
\eqref{xilowerbound}, we obtain the estimate
\begin{equation}
\text{III}\geq e^{-\eps (t-\gamma_t)\alpha_{b_t}^{-2}}\,
\E_0\Bigl[\exp\Bigl\{\int_0^{t-\gamma_t}\psi_{t}\bigl(X(s)\bigr)\,
ds\Bigr\} \1\{\tau_{0,t}>t-\gamma_t\}\Bigr],
\end{equation}
By invoking \eqref{FKDir} and \eqref{Fourieru}, the expectation on
the right-hand side is bounded from below by
\begin{equation}
\exp\left\{(t-\gamma_t)\lambda^{\rm d}(t)
\right\} {\rm e}_t(0)^2,
\end{equation}
where $\lambda^{\rm d}(t)$ resp.\ ${\rm e}_t$ denote the principal
Dirichlet eigenvalue resp.\ the $\ell^2$-normalized principal
eigenfunction of $\kappa\DeltaD+\psi_{t}$ in $Q^{(t)}$.
For $e_t(0)$ and $\lambda^{\rm d}(t)$ we
have the following bounds, whose proofs will be given
subsequently:

\begin{lemma}
\label{aux}
We have
\begin{align}\label{vectappro}
\ &\liminf_{t\to\infty}\frac{\alpha_{b_t}^2}t\log {\rm
e}_t(0)^2\geq 0,\\
\label{eigenappro}
\ &\liminf_{t\to\infty}\alpha_{b_t}^2\lambda^{\rm d}(t)
\ge\lambda_R(\psi).
\end{align}
\end{lemma}

Summarizing all the preceding estimates and applying
\eqref{vectappro} and \eqref{eigenappro}, we obtain
\begin{equation}
\liminf_{t\to\infty}\frac{\alpha_{b_t}^2}t\log u(t,0)\geq
\lambda_R(\psi)-\eps,
\end{equation}
where we also noted that $t-\gamma_t=t(1+o(1))$. In the case
$\gamma>0$, let $\eps\downarrow 0$, optimize over $\psi\in C^-(R)$
with ${\mathcal L}_R(\psi)<d$ (clearly, the supremum in
\eqref{chitildedef} may be restricted to the set of twice
continuously differentiable functions $\psi\in C^{-}(R)$ such that
${\mathcal L}_R(\psi)<d$) and let $R\to\infty$ to get the lower
bound in Theorem~\ref{asasy}. In the case $\gamma=0$, recall that
${\mathcal L}_R(\psi)=\const |\{\psi<0\}|$. It is classical (see,
e.g., \cite{DV75}, Lemma~3.13, or argue directly by Faber-Krahn's
inequality) that the supremum \eqref{chitildedef} can be
restricted to $\psi$ whose support is a ball. The proof is
therefore finished by letting $\eps\downarrow0$, optimizing over
such $\psi$ and letting $R\to\infty$.
\end{proofsect}
\begin{proofsect}{Proof of Lemma~\ref{aux}}
We begin with \eqref{vectappro}. Recall that ${\rm e}_t$ is also
an eigenfunction for the transition densities of the random walk
in $Q^{(t)}$ with potential ${\psi}_{t}-\lambda^{\rm d}(t)$. Using
this observation at time $1$, we can write
\begin{equation}
{\rm e}_t(0)=\E_0\Bigl[\exp\Bigl\{\int_0^1
\bigl[\psi_{t}\bigl(X(s)\bigr)-\lambda^{\rm d}(t)\bigr]\,ds\Bigr\}
\1\{\tau_{0,t}>1\}{\rm e}_t\bigl(X(1)\bigr)\Bigr],
\end{equation}
Since $\lambda^{\rm d}(t)$ is nonpositive and $\psi$ is bounded
from below, we have
\begin{equation}
{\rm e}_t(0)\geq \exp\bigl[\alpha(b_t)^{-2}\inf \psi\bigr]
\sum_{z\in Q^{(t)}}
\P_0\bigl(\tau_{0,t}>1,X(1)=z\bigr){\rm e}_t(z).
\end{equation}
Using the same strategy as in \eqref{prob1}, we have
$\P_0(\tau_{0,t}>1,X(1)=z)\geq
e^{-O(\alpha(b_t)\log\alpha(b_t))}$. Since ${\rm e}_t$ is
nonnegative and satisfies $\|{\rm e}_t\|_2=1,$ we have $\sum_z{\rm
e}_t(z)\geq \|{\rm e}_t\|_2^2=1$. From these estimates,
\eqref{vectappro} is proved by noting that
$\alpha(b_t)\log\alpha(b_t))=o(t/\alpha(b_t)^2)$.

In order to establish \eqref{eigenappro}, we shall restrict the
supremum in \eqref{discreteeigenv} to a particular choice of~$g$.
Let $Q_R(\psi)=[-R,R]^d$ if $\gamma\not=0$ and
$Q_R(\psi)=\supp\psi$ if $\gamma=0$. Let $\widehat g\colon
[-R,R]^d\to[0,\infty)$ be the $L^2$-normalized principal
eigenfunction of the (continuous) operator $\kappa\Delta+\psi$ on
$Q_R(\psi)$ with Dirichlet boundary conditions. Let us insert
$\widehat g_t(z)=\widehat g(z/\alpha(b_t))/\alpha(b_t)^{d/2}$ into
\eqref{discreteeigenv} in the place of $g$. Thus we get
\begin{equation}
\label{bd}
\alpha(b_t)^2\lambda^{\rm d}(t)(\psi_{t})\geq
\alpha(b_t)^{-d}\!\!\!\sum_{z\in Q^{(t)}} \Bigl[(\psi \widehat
g^2)\bigl({\textstyle\frac z{\alpha(b_t)} }
\bigr)-\kappa\alpha(b_t)^2\sum_{y\colon y\sim z}\Bigl(\widehat
g\bigl( \textstyle\frac z{\alpha(b_t)} \bigr)-\widehat g\bigl(
\textstyle\frac {y}{\alpha(b_t)} \bigr)\Bigr)^2\Bigr],
\end{equation}
where $y\sim z$ denotes that $y$ and $z$ are nearest neighbors.

Since $\psi$ is smooth, standard theorems guarantee  that
$\widehat g$ is continuously differentiable on $Q_R(\psi)$ and,
hence, $\Vert \nabla \widehat g\Vert_\infty<\infty$. (This fact is
derived using regularity properties of Green's function of the
Poisson equation, see, e.g., Theorem~10.3 in Lieb and Loss
\cite{LL97}.) Then
\begin{equation}
\widehat g\bigl(z/\alpha(b_t)\bigr)-\widehat
g\bigl(y/\alpha(b_t)\bigr)=\alpha(b_t)^{-1}(z-y)\cdot\nabla \widehat
g\bigl(z_\eta/\alpha(b_t)\bigr), \qquad z,y\in Q^{(t)},
\end{equation}
where $z_\eta=\eta z +(1-\eta) y$ for some $\eta\in[0,1]$. For the
pairs $z\sim y$ with $y\not\in Q^{(t)}$ we only get a
bound $|\widehat g(z/\alpha(b_t))-\widehat g(y/\alpha(b_t))|\le
(1+\Vert \nabla \widehat g\Vert_\infty)/\alpha(b_t)$ (note that
$\widehat g(y/\alpha(b_t))=0$ in this case). Since the total
contribution of these boundary terms to \eqref{bd} is clearly
bounded by $(1+\Vert \nabla \widehat g\Vert_\infty)/\alpha(b_t)$,
we see that the right-hand side of \eqref{bd} converges to
$(\psi,\widehat g^2)-\kappa\|\nabla \widehat g\|_2$ as
$t\to\infty$. By our choice of $\widehat g$, this limit is equal
to the eigenvalue $\lambda_R(\psi)$, which proves
\eqref{eigenappro}. \end{proofsect}

\begin{proofsect}{Proof of Theorem~\ref{asasy} ($d=1$), lower bound}
Suppose that $\langle\log(-\xi(0)\vee1)\rangle>-\infty$. This
implies that ${\mathcal C}_\infty=\Z$ almost surely and, by the
law of large numbers,
\begin{equation}
\label{sumofxi}
K_\xi:=\sup_{y\in \Z\setminus\{0\}}\frac1{|y|}
\sum_{x=0}^{|y|}\log\bigl(-\xi(x)\vee1\bigr)<\infty
\quad \text{almost surely.}
\end{equation}
Suppose that $\xi=(\xi(z))_{z\in\Z}$ does not belong to the exceptional
sets of \eqref{sumofxi} and Proposition~\ref{ximicrobox}. For
sufficiently large $t$, let $y_t\in Q_{\gamma_t}$ be such
that \eqref{xilowerbound} holds.

Let $r_x=(-1/\xi(x))\wedge1$. The strategy for the lower bound on
$u(t,0)$ is that the random walk performs $|y_t|$ steps toward
$y_t$, resting at most time $r_x$ at each site $x$ between $0$ and
$y_t$, so that $y_t$ is reached before time $\gamma_t$. Afterwards
the walk stays at $y_t$ until $\gamma_t$. Use $E^{(t)}$ to
denote the latter event. Then $u(t,0)\geq
\text{II}\times\text{III}$, where $\text{III}$ is as in
\eqref{IIIIII} and
$\text{II}=\E_0\bigl[e^{\int_0^{\gamma_t}\xi(X(s))\,ds}\1_{E^{(t)}}\bigr]$.

The lower bound on $\text{III}$ is identical to the case $d\ge2$.
To estimate the term $\text{II}$, suppose that $y_t>0$ (clearly,
if $y_t=0$ no estimate on $\text{II}$ is needed; $y_t<0$ is handled
by symmetry) and abbreviate $|y_t|=n+1$. Using the shorthand
$[s]_n=s_0+\dots+s_n$, we have
\begin{multline}
\label{IIford=1}
\quad\text{II}= \int_0^{r_0}\!ds_0\,\dots\int_0^{r_n}\!ds_n\,
\int_0^{\gamma_t-[s]_n}
\!ds_{n+1}\,\exp\Bigl\{-\sum_{x=0}^{n+1} s_x\bigl(\kappa-\xi(x)\bigr)\Bigr\}
\\
\ge
e^{O(\gamma_t)}\prod_{x=0}^n\Bigl[r_x\exp\bigl(r_x\xi(x)\bigr)\Bigr]
\ge e^{O(\gamma_t)}
\exp\Bigl\{-\sum_{x=0}^n\log\bigl(-\xi(x)\vee1\bigr)\Bigr\}.
\end{multline}
Indeed, in the first line we noted that $[s]_n\le\gamma_t$ because
$r_x\le 1$. Then we took out the terms $\exp(-\kappa s_x)$ as well
as $\exp(s_{n+1}\xi(y_t))$, recalling that
$\xi(y_t)\ge\inf\psi_t=\inf\psi/\alpha(b_t)^2=O(1)$ and that
$|y_t|=O(\gamma_t)$. The last inequality follows by the fact that
$r_x\exp(r_x\xi_x)\ge r_x/e$. Invoking \eqref{sumofxi}, the sum in
the exponent is bounded above by $K_\xi|y_t|=O(\gamma_t)$, whereby
we finally get that $\text{II}\ge e^{-O(\gamma_t)}$. \end{proofsect}
\subsection{Technical claims\label{tech-claims}}
In this final subsection, we prove Proposition~\ref{ximicrobox}. First, 
we need to
introduce some notation and prove two auxiliary lemmas.
For $\eps>0$ and $y\in\Z^d$, define the event
\begin{equation}\label{Adef}
A_y^{(t)}=\{y\in {\mathcal
C}_\infty^*\}\cap\bigcap_{z\in Q^{(t)}}\Bigl\{\xi(y+z)\geq
\psi_{t}(z)\textstyle{-\frac \eps {2\alpha(b_{t})^2}}\Bigr\}.
\end{equation}
Note that the distribution of $A_y^{(t)}$ does not depend on $y$.
By $\partial(Q)$ we denote the outer boundary of a set
$Q\subset\Z^d$. To estimate $\prob(A_y^{(t)})$, it is convenient
to begin with the first event on the right-hand side of
\eqref{Adef}. Since $\{y\in{\mathcal C}_\infty^*\}\subset
\partial (y+Q^{(t)})\cap {\mathcal C}_\infty^*$ it suffices to know
an estimate on $\prob(\partial Q^{(t)}\cap {\mathcal C}_\infty^*)$:

\begin{lemma}\label{probB}
Let $d\ge2$ and let $\psi\in C^-(R)$ be such that $\psi\not\equiv0$. Then
there is a $c\in(0,\infty)$ such that, for $t$ large enough,
\begin{equation}\label{Bprobesti}
\prob\bigl(\partial Q^{(t)}\cap {\mathcal C}_\infty^*
=\emptyset\bigr)\leq e^{-c\alpha(b_t)} .
\end{equation}
\end{lemma}
\vspace{0.1cm}

\begin{proofsect}{Proof}
Since $\psi\not\equiv0$ is continuous, there is a ball
$B_{\alpha(b_t)}$ of radius of order $\alpha(b_t)$ such that
$B_{\alpha(b_t)}\subset Q^{(t)}$. If $t$ is so large that
$\psi_t\ge\inf\psi/\alpha(b_t)^2\ge-K$, then
$B_{\alpha(b_t)}\subset\{z\colon\xi(z)\ge-K\}$ and the left-hand
side of \eqref{Bprobesti} is bounded from above by $\prob(\partial
B_{\alpha(b_t)}\cap {\mathcal C}_\infty^*=\emptyset)$. The proof
now proceeds in a different way depending whether $d\ge3$ or
$d=2$. In the following, the words ``percolation,'' ``infinite
cluster,'' etc., refer to site-percolation on $\Z^d$ with
parameter $p=\prob(\xi(0)>-K)$. Recall that $p>p_{\rm c}(d)$ by
our choice of $K$.

Let $d\ge3$. Then, by equality of $p_{\rm c}(d)$ and the limit of
slab-percolation thresholds, there is a width $k$ such that the
slab $S_k=\Z^{d-1}\times\{1,\dots,k\}$ contains almost surely an
infinite cluster. Pick a lattice direction and decompose $\Z^d$
into a disjoint union of translates of $S_k$. There is $c'>0$ such
that, for $t$ large, at least $\lfloor c'\alpha(b_t)/k\rfloor$
slabs are intersected by $\partial B_{\alpha(b_t)}$. Then
$\{\partial B_{\alpha(b_t)}\cap {\mathcal C}_\infty^*=\emptyset\}$
is contained in the event that in none of the slabs intersecting
$\partial B_{\alpha(b_t)}$ the respective infinite cluster reaches
$\partial B_{\alpha(b_t)}$. Let $P_\infty(k)$ be minimum
probability that a site in $S_k$ belongs to an infinite cluster.
Combining the preceding inclusions, we have
\begin{equation}
\prob(\partial B_{\alpha(b_t)}\cap {\mathcal
C}_\infty^*=\emptyset)\le P_\infty(k)^{c'\alpha(b_t)/k}.
\end{equation}
Now the claim follows by putting $c=-c'k^{-1}\log P_\infty(k)$.

In $d=2$, suppose without loss of generality that
$B_{\alpha(b_t)}$ is centered at the origin. Recall that $x$ and
$y$ are $*$-connected if their Euclidean distance is not more than
$\sqrt2$. On the event $\{\partial B_{\alpha(b_t)}\cap {\mathcal
C}_\infty^*=\emptyset\}$, the origin is encircled by a
$*$-connected circuit of size at least $c\alpha(b_t)$ for some
$c>0$, not depending on $t$. Denote by $x$ the nearest point of
this circuit in the first coordinate direction. Call sites $z$
with $\xi(z)\ge-K$ ``occupied,'' the other sites are ``vacant.''

Note that percolation of occupied sites rules out percolation of
vacant sites, e.g., by the result of Gandolfi, Keane, and
Russo~\cite{GKR88}. Moreover, using the site-perolation version of
the famous ``$p_c=\pi_c$'' result (see e.g., Grimmett~\cite{G89}),
the probability that a given site is contained in a vacant
$*$-cluster of size $n$ is bounded by $e^{-\sigma(p)n}$, where
$\sigma(p)>0$ since $p>p_c(d)$. If the ball $B_{\alpha(b_t)}$ has
diameter at least $r\alpha(b_t)$, then by taking the above circuit for
such a cluster we can estimate the probability of its occurrence:
\begin{equation}
\prob\bigl(\partial Q^{(t)}\cap {\mathcal
C}_\infty^*=\emptyset\bigr) \le\sum_{n=\lfloor
r\alpha(b_t)\rfloor}^\infty n e^{-\sigma(p)n}\le
e^{-\sigma(p)r\alpha(b_t)/2},
\end{equation}
for $t$ large enough. Here ``$n$'' in the sum accounts for the
position of the circuit's intersection with the positive part of
the first coordinate axis. The minimal size of the circuit is at
least $\lfloor r\alpha(b_t)\rfloor$, since it has to stay all
outside $B_{\alpha(b_t)}$. The claim follows by putting
$c=r\sigma(p)/2$.
\end{proofsect}

\begin{lemma}\label{probA} For any $\eps>0$,
\begin{equation}\label{Aprobesti}
\prob(A_0^{(t)})\geq t^{-{\mathcal L}_R(\psi)+o(1)},\qquad t\to\infty.
\end{equation}
\end{lemma}
\vspace{0.1cm}

Let $H$ be in the $\gamma$-class and let $\psi\not\equiv0$ 
(otherwise there is nothing to prove because ${\mathcal 
L}_R(0)=\infty$). Consider the event
\begin{equation}
\label{tildeAt}
\widetilde A^{(t)}=\bigcap_{z\in Q^{(t)}}\Bigl\{\xi(z)\geq
\psi_t(z)\textstyle{-\frac \eps {2\alpha(b_t)^2}}\Bigr\}.
\end{equation}
Note that both events on the right-hand side of \eqref{Adef} are
increasing in the partial order $\xi\succeq\xi'$ $\Leftrightarrow$
$\xi(x)\ge\xi'(x)$ for all $x$. Therefore, by the FKG-inequality,
\begin{equation}
\label{5.36}
\prob(A_0^{(t)})\ge \prob(0\in{\mathcal 
C}_\infty^*)\,\prob(\widetilde A^{(t)}).
\end{equation}
Since $\prob(0\in{\mathcal C}_\infty^*)>0$, we only need to prove 
the assertion for $A_0^{(t)}$ replaced by $\widetilde A^{(t)}$. 
The proof proceeds in three steps, depending on $\gamma$ and on 
whether there is an atom at $0$.

\begin{proofsect}{Proof of Lemma~\ref{probA} for $\gamma\in(0,1)$}
Let $f\in C^+(R)$ be the solution to $\psi-\frac 38 \eps=\H'\circ
f$ and let $f_t\colon\Z^d\to(0,\infty)$ be its scaled version:
$f_t(z)=(b_t/\alpha(b_t)^d)f(z/\alpha(b_t))$. Define the tilted
probability measure
\begin{equation}
\prob_{t,z}(\bdot)=\bigl\langle
e^{f_t(z)\xi(z)}\1\{\xi(z)\in \bdot\}
\bigr\rangle e^{-H(f_t(z))}.
\end{equation}
We denote expectation with respect to
$\prob_{t,z}$ by $\langle\bdot\rangle_{t,z}$. Consider the event
\begin{equation}
\label{D_t}
D_t(z)=\Bigl\{-\frac\eps{4\alpha(b_t)^2}\geq\xi(z)-\psi_t(z)\geq
-\frac\eps{2\alpha(b_t)^2}\Bigr\}.
\end{equation}
Then $\prob(\widetilde A^{(t)})$ can be bounded as
\begin{equation}
\prob\bigl(\widetilde A^{(t)}\bigr)\geq\prod_{z\in
Q^{(t)}}\Bigl[e^{H(f_t(z))} \bigl\langle
e^{-f_t(z)\xi(z)}\1\{D_t(z)\}\bigr\rangle_{t,z}\Bigr].
\end{equation}
Applying the left inequality in \eqref{D_t}, we obtain
\begin{equation}
\prob\bigl(\widetilde A^{(t)}\bigr)\ge\exp\Bigl\{\sum_{z\in
Q^{(t)}}\bigl[{\textstyle{H(f_t(z))- f_t(z)\bigl(\psi_t(z)- \frac
\eps{4\alpha(b_t)^2}\bigr)}}\bigr]\Bigr\} \prod_{z\in
Q^{(t)}}\prob_{t,z}\bigl(D_t(z)\bigr).
\end{equation}
Since $\gamma>0$ and $f$ is continuous and bounded, we can use our
Scaling Assumption and the fact that $b_t\alpha(b_t)^{-2}=\log t$
to turn the sum over $z\in Q^{(t)}$ into a Riemann integral over
$[-R,R]^d$:
\begin{equation}
\prob(\widetilde A^{(t)})\geq   t^{-\int[f\psi-\widetilde H\circ
f] +\frac\eps 4\int f+o(1)}\prod_{z\in
Q^{(t)}}\prob_{t,z}\bigl(D_t(z)\bigr).
\end{equation}
where we also used that $Q^{(t)}=Q_{R\alpha(b_t)}$ in this case.
In order to finish the proof of the lower bound in
\eqref{Aprobesti}, we thus need to show that
\begin{equation}\label{fbound}
\int \bigl[f\psi-\widetilde H\circ
f\bigr]\leq {\mathcal L}_R(\psi),
\end{equation}
and that
\begin{equation}\label{LLN}
\prod_{z\in Q^{(t)}}\prob_{t,z}\bigl(D_t(z)\bigr)\ge
t^{o(1)},\qquad t\to\infty.
\end{equation}

Let us begin with \eqref{fbound}. For simplicity, we restrict
ourselves to the case when $\H(1)=-1$. Then
${\mathcal L}_R(\psi)=\gamma^{1/(1-\gamma)}(\gamma^{-1}-1)
\int|\psi|^{-\gamma/(1-\gamma)}$ and $f=\gamma^{1/(1-\gamma)}
|\psi-\frac38\eps|^{-1/(1-\gamma)}$. Hence,
\begin{equation}
 \int \bigl[f\psi-\widetilde H\circ
f\bigr]-{\mathcal L}_R(\psi)=\gamma^{\frac
1{1-\gamma}}\int|\psi|^{-\frac
\gamma{1-\gamma}}\zeta_\gamma\Bigl(
\textstyle{\bigl|\frac\psi{\psi-\frac
38\eps}\bigr|^{\frac1{1-\gamma}}}\Bigr),
\end{equation}
where $\zeta_\gamma(x)=1-x-\frac 1\gamma (1-x^\gamma)$. Since
$\zeta_\gamma(x)\leq 0$ for any $x\ge0$, \eqref{fbound} is proved.

In order to prove \eqref{LLN}, note that
\begin{multline}
\label{LLNproof1}
\quad\prob_{t,z}\bigl(D_t(z)\bigr)\geq 1-
\prob_{t,z}
\Bigl(\xi(z)\geq \psi_t(z)-\frac \eps{4\alpha(b_t)^2}\Bigr)\\
-\prob_{t,z}
\Bigl(\xi(z)\leq \psi_t(z)-\frac \eps{2\alpha(b_t)^2}\Bigr).
\quad
\end{multline}
We concentrate on estimating the second term; the first term is
handled analogously. By the exponential Chebyshev inequality, we
have for any $g_t(z)\in(0,f_t(z))$ that
\begin{equation}
\label{est}
\begin{aligned}
\prob_{t,z}&\Bigl(\xi(z)\leq\psi_t(z) -\frac\eps
{2\alpha(b_t)^2} \Bigr)\\ &\leq e^{-H(f_t(z))}\Bigl\langle
\exp\Bigl\{f_t(z)\xi(z)-g_t(z)
\bigl[\xi(z)-\psi_t(z)+\textstyle\frac\eps{2\alpha(b_t)^2}\bigr]\Bigr\}
\Bigr\rangle \\
&=\exp\Bigl\{{H\bigl(f_t(z)-g_t(z)\bigr)-H\bigl(f_t(z)\bigr)
+g_t(z)\psi_t(z)-g_t(z)\frac\eps{2\alpha(b_t)^2}}\Bigr\}.
\end{aligned}
\end{equation}

Note that $\H_t'\to \H'$ (recall \eqref{H_t}) as $t\to\infty$
uniformly on compact sets in $(0,\infty)$. Also note that $f$ is
bounded away from $0$. Choose $g_t(z)=\delta_t f_t(z)$, where
$\delta_t\downarrow 0$ is still to be chosen appropriately. Then
the exponent in the third line of \eqref{est} can be bounded from
above by
\begin{multline}
-\delta_t\frac{b_t}{\alpha(b_t)^{d+2}}
f\Bigl(\frac{z}{\alpha(b_t)}\Bigr)\biggl\{
\H_t'\Bigl[f\Bigl(\frac{z}{\alpha(b_t)}\Bigr)(1-\delta_t)\Bigr]
-\psi\Bigl(\frac{z}{\alpha(b_t)}\Bigr)+\frac\eps2\biggr\}\\
=-\delta_t\frac{b_t}{\alpha(b_t)^{d+2}}f\Bigl(
\frac{z}{\alpha(b_t)}\Bigr)\Bigl[\frac\eps 8+o(1)\Bigr],
\end{multline}
where we replaced $\widetilde H_t'$ by $\widetilde H'+o(1)$ and
used the definition relation for $f$. Pick
$\delta_t=(\alpha_{b_t}^{d+2}/b_t)^{1/2}$ for definiteness.
Taking the product over $z\in Q^{(t)}$ in \eqref{LLNproof1} and
using that $[\frac\eps8+o(1)]f\ge C>0$, we obtain for $t$ large
that
\begin{multline}
\prod_{z\in Q^{(t)}}\prob_{t,z}\bigl(D_t(z)\bigr)\geq
\Bigl[1-2\exp\Bigl\{-C\delta_t
\frac{b_t}{\alpha(b_t)^{d+2}}\Bigr\}\Bigr]^{\#Q^{(t)}}\\
\qquad\geq\exp\Bigl\{-4\#Q^{(t)}\exp\Bigl\{-C{\delta_t\frac{b_t}
{\alpha(b_t)^{d+2}}}\Bigr\} \Bigr\}=t^{-C'
(\alpha_{b_t}^{d+2}/b_t)\exp\bigl(-C
\delta_t\frac{b_t}{\alpha(b_t)^{d+2}}\bigr)},
\end{multline}
where also used that $b_t\alpha(b_t)^{-2}=\log t$ and
$\#Q^{(t)}\le\alpha(b_t)^dC'/4$ for some $C'$ as $t\to\infty$. By
our choice of $\delta_t$, \eqref{LLN} is clearly satisfied, which
finishes the proof in the case $\gamma\in(0,1)$.
\end{proofsect}

\begin{proofsect}{Proof of Lemma~\ref{probA} for $\gamma=0$, atom at $0$}
Suppose $\prob(\xi(0)\in\cdot)$ has an atom at $0$ with mass
$p>0$. Then, noting that $Q^{(t)}$ are only the sites with
$\psi_t<0$, we have
\begin{equation}
\prob(\widetilde A^{(t)})\geq
\prob\bigl(\xi(0)=0\bigr)^{\#Q^{(t)}}=\exp\bigl\{
\alpha(b_t)^d(|\supp\psi|+o(1))\log p\bigr\},\quad t\to\infty.
\end{equation}
Since $\alpha_t=t^{1/(d+2)}$ and $\widetilde H(1)=\log p$, we have
${\mathcal L}_R(\psi)=-\widetilde H(1)|\supp\psi|$ and
$\alpha(b_t)^d=\log t$, whereby
\eqref{Aprobesti} immediately follows.
\end{proofsect}

\begin{proofsect}{Proof of Lemma~\ref{probA} for $\gamma=0$, no atom at $0$}
Suppose that $\gamma=0$ and $\prob(\xi(0)=0)=0$. Set
$f_t=b_t\alpha(b_t)^{-d}$ and consider the probability measure
$\prob_t(\xi(0)\in\cdot)$ with density $\exp[f_t\xi(0)-H(f_t)]$
with respect to $\prob(\xi(0)\in\cdot)$. Invoking that $\xi(0)\leq
0$, we obtain
\begin{equation}
\prob(\widetilde A^{(t)})\geq \prob\Bigl(\xi(0)\geq {
-\frac \eps{2\alpha(b_t)^2}}\Bigr)^{\#Q^{(t)}}
\geq e^{\#Q^{(t)}H(f_t)}\,\prob_t\Bigl(\xi(0)\ge{
-\frac \eps{2\alpha(b_t)^2}}\Bigr)^{\#Q^{(t)}}.
\end{equation}
Now use the Scaling Assumption and the fact that 
$\#Q^{(t)}=\alpha(b_t)^d(|\supp\psi|+o(1))$ as $t\to\infty$ to 
extract the term $t^{-{\mathcal L}_R(\psi)}$ from the exponential 
on the right-hand side (here we recalled that ${\mathcal 
L}_R(\psi)=-\widetilde H(1)|\supp\psi|$). Moreover, by an 
argument similar to \eqref{est}, the last term on the right-hand 
side is no smaller than $t^{o(1)}$ as $t\to\infty$. To that end 
we noted that our choice of $f_t$ corresponds to $f\equiv1$ and 
then we used again that 
$\lim_{t\to\infty}b_t\alpha(b_t)^{-(d+2)}=\infty$, which follows 
from the fact that $\xi(0)$ has no atom at zero. This finally 
finishes the proof of Lemma~\ref{probA}. \end{proofsect}
Now we can finish off the proof of Proposition~\ref{ximicrobox}.

\begin{proofsect}{Proof of Proposition~\ref{ximicrobox}}
Fix $R>0$ and $\psi\in C^-(R)$ with ${\mathcal L}_R(\psi)< d$.
Recall the notation \eqref{Qdef} and \eqref{Adef}. Let
$t_1=t_1(\psi,\eps,R)$ be such that for all $t\ge t_1$ and for all
$s\in[0,e)$
\begin{equation}
\label{unif-cont}
\psi_{et}(z)-\frac\eps{2\alpha(b_{et})^2}\geq\psi_{st}(z)
-\frac\eps{\alpha(b_{st})^2},\qquad z\in Q^{(st)}.
\end{equation}
Such a $t_1<\infty$ indeed exists, since
$\alpha(b_{st})/\alpha(b_{et})\to1$ as $t\to\infty$ and since
$\psi$ is uniformly continuous on
$[-R,R]^d$. This implies that to prove Proposition~\ref{ximicrobox}
it suffices to find an almost-surely finite
$n_0=n_0(\xi,\psi,\eps,R)$ such that for each $n\ge n_0$ there is
a $y_n\in Q_{\gamma_{e^n}}$ for which the event
$A_{y_n}^{(e^{n+1})}$ occurs. Indeed, for any $t=s e^n$ with $n\ge
n_0$ and $s\in[0,e)$ we have that
$Q_{\gamma_{e^n}}\subset Q_{\gamma_t}$ and
$y_n+Q_{R\alpha(b_t)}\subset y_n+Q_{R\alpha(b_{e^{n+1}})}$,
as follows by monotonicity of the maps $t\mapsto\gamma_t$ and
$t\mapsto\alpha(b_t)$ and, consequently,
\begin{equation}
\bigcap_{z\in Q^{(t)}}\bigl\{\xi(y_n+z)\ge\psi_t(z)
-\textstyle\frac\eps{\alpha(b_t)^2}\bigr\}\supset A_{y_n}^{(e^{n+1})},
\end{equation}
by invoking \eqref{unif-cont}. Then Proposition~\ref{ximicrobox}
would follow with the choice $t_0=t_1\vee e^{n_0}$.

Based on the preceding reduction argument, let $t\in\{e^n\colon
n\in\N\}$ for the remainder of the proof.  Let
$M_t=Q_{\gamma_{t}}\cap \lfloor 3R\alpha(b_{et})\rfloor\Z^d$. We
claim that, to prove Proposition~\ref{ximicrobox} for
$t\in\{e^n\colon n\in\N\}$, it suffices to show the summability of
\begin{equation}
p_t=\prob\Bigl(\sum_{y\in M_t}\1_{A_y^{(et)}}\leq \textstyle\frac 12\#M_t\,
\prob\bigl(A_0^{(et)}\bigr)\Bigr),\qquad t\in\{e^n\colon n\in\N\}.
\end{equation}
Indeed, since $\#M_t\geq t^{d+o(1)}$ we have by Lemma~\ref{probA}
\begin{equation}\label{probesti}
\#M_t\,\prob(A_0^{(et)})\geq t^{d-{\mathcal
L}_R(\psi)+o(1)},\qquad t\to\infty.
\end{equation}
Since we assumed ${\mathcal L}_R(\psi)< d$, summability of $p_t$
would imply the existence of at least one site $y\in Q_{\gamma_t}$
(in fact, at least $t^{d-{\mathcal L}_R(\psi)+o(1)}$ sites) with
$A_y^{(et)}$ satisfied.

To prove a suitable bound on $p_t$ we invoke
Chebyshev's inequality to find that
\begin{equation}\label{pesti1}
p_t\leq \frac 4{\#M_t\,\prob(A_0^{(et)})}+
\frac{4\max_{y\not= y'}\mbox{cov}(A_y^{(et)},
A_{y'}^{(et)})}{\prob(A_0^{(et)})^2}.
\end{equation}
As follows from \eqref{probesti}, the first term on the right-hand
side is summable on $t\in\{e^n\colon n\in\N\}$. In order to
estimate $\mbox{cov}(A_y^{(et)},A_{y'}^{(et)})$ for $y\not=y'$,
let $\HH$ and $\HH'$ be two disjoint half spaces in ${\mathbb
R}^d$ which contain $y+Q^{(et)}$ and $y'+Q^{(et)}$, respectively,
including the outer boundaries. By our choice of $M_t$, $\HH$ can
be chosen such that $\dist(y+Q^{(et)},{\mathbb
H}^{\text{c}})\ge R\alpha(b_t)/3$, and similarly for $\HH'$.
We introduce the event $F_y$ that the outer boundary of $y+Q^{(et)}$
is connected to infinity by a path in ${\mathcal
C}_\infty^*\cap\HH$, and the analogous event $F_{y'}$ with $y'$
and $\HH'$ instead of $y$ and $\HH$. By splitting $A_y^{(et)}$
into $A_y^{(et)}\cap F_y$ and $A_y^{(et)}\cap F_y^{\rm c}$ (and
analogously for $y'$) and invoking the independence of
$A_y^{(et)}\cap F_y$ and $A_{y'}^{(et)}\cap F_{y'}$  we see that
\begin{equation}
\begin{aligned}
\mbox{cov}\bigl(A_y^{(et)},A_{y'}^{(et)}\bigr)&=
\mbox{cov}\bigl(A_y^{(et)}\cap F_y^{\text{c}},
A_{y'}^{(et)}\bigr)+\mbox{cov}\bigl(A_y^{(et)}\cap F_y,
A_{y'}^{(et)}\cap F_{y'}^{\text{c}}\bigr)\\
&\le \prob\bigl(\widetilde A^{(et)}\bigr)^2\bigl[\prob(F_y^{\text{c}})
+\prob(F_{y'}^{\text{c}})\bigr],
\end{aligned}
\end{equation}
where we recalled \eqref{tildeAt} for the definition of
$\widetilde A^{(et)}$.

In order to estimate the last expression, let us observe that
\begin{equation}
F_y^{\text{c}}\subset\bigl\{\partial(y+Q^{(et)})\cap{\mathcal
C}_\infty^*=\emptyset\bigr\}\cup \bigcup_{x\in
\partial(y+Q^{(et)})}G_x
\end{equation}
where $G_x$ is the event that $x$ is in a finite component of
$\{z\colon\xi(z)\ge-K\}\cap {\mathbb H}$ which reaches up to
${\mathbb H}^{\text{c}}$. By Lemma~\ref{probB}, the probability of
the first event is bounded by $e^{-c\alpha(b_t)/2}$ and, as is
well known (see, e.g., Grimmett~\cite{G89}, proof of
Theorem~6.51), $\prob(G_x)$ is exponentially small in
$\text{dist}(x,{\mathbb H}^{\text{c}})$, which is at
least $R\alpha(b_t)/3$. Since
$\#\partial(y+Q^{(et)})=O(\alpha(b_t)^{d-1})$, we have
\begin{equation}
\prob(F_y^{\text{c}})\le e^{-c_*\alpha(b_t)}
\end{equation}
for some $c_*>0$. Since $\alpha(b_t)=n^{\nu/(1-2\nu)+o(1)}$ for 
$t=e^n$, also the second term is thus summable on $t\in\{e^n\colon 
n\in\N\}$, because by \eqref{5.36}, $\prob(\widetilde 
A^{(et)})\le \prob(A^{(et)})/\prob(0\in{\mathcal C}_\infty^*)$. 
Combining all the preceding reasoning, the proof of 
Proposition~\ref{ximicrobox} is finished. 
\end{proofsect}

\vbox{
\section*{Acknowledgments}
\noindent
\nopagebreak W.K.\ would like to thank J\"urgen G\"artner for
various occasional discussions. The authors wish to acknowledge
the hospitality of TU Berlin (M.B.) and Microsoft Research
(W.K.). 

}


\end{document}